\documentclass[aps,pre,twocolumn,showpacs,superscriptaddress,preprintnumbers,amsmath,amssymb,floatfix,10pt]
{revtex4-1}
\newif\ifbw
\bwfalse

\usepackage{graphicx}
\usepackage{verbatim}
\usepackage{cancel}

\newcommand{\beq}{\begin{equation}}
\newcommand{\eeq}{\end{equation}}

\newcommand{\beqa}{\begin{eqnarray}}
\newcommand{\eeqa}{\end{eqnarray}}

\newcommand \rt {\right}
\newcommand \lt {\left}
\newcommand \nline {\nonumber \\}

\newcommand{\mbfk}{{\mathbf k}}
\newcommand{\mbfq}{{\mathbf q}}
\newcommand{\mbfr}{{\mathbf r}}
\newcommand{\mbfi}{{\mathbf i}}
\newcommand{\mbfK}{{\mathbf K}}

\begin{document}
\title{Complex order-parameter phase-field models derived from structural phase-field-crystal models}

\author{Nana Ofori-Opoku}
\affiliation{Department of Materials Science and Engineering and
Brockhouse Institute for Materials Research,
McMaster University, Hamilton, Canada L8S-4L7}
\affiliation{Department of Physics and Centre for the Physics of Materials, Rutherford Building, McGill University, Montreal, Canada, H3A-2T8}

\author{Jonathan Stolle}
\affiliation{Department of Physics and Astronomy and
Brockhouse Institute for Materials Research,
McMaster University, Hamilton, Canada L8S-4M1 }

\author{Zhi-Feng Huang}
\affiliation{Department of Physics and Astronomy, Wayne State University, Detroit, USA, 48201}

\author{Nikolas Provatas}
\affiliation{Department of Physics and Centre for the Physics of Materials, Rutherford Building, McGill University, Montreal, Canada, H3A-2T8}

\begin{abstract}
The phase-field-crystal (PFC) modeling paradigm is rapidly emerging as the model of choice when investigating materials phenomena with atomistic scale effects over diffusive time scales. Recent variants of the PFC model, so-called structural PFC (XPFC) models introduced by Greenwood {\it et al.}, have further increased the capability of the method by allowing for easy access to various structural transformations in pure materials [Phys. Rev. Lett. {\bf 105}, 045702 (2010)] and binary alloys [Phys. Rev. B. {\bf 84}, 064104, (2011)]. We present an amplitude expansion of these XPFC models, leading to a mesoscale complex order-parameter (amplitude), i.e., phase-field representation, model for two dimensional square-triangular structures. Amplitude models retain the salient atomic scale features of the underlying PFC models, while resolving microstructures on mesoscales as in traditional phase-field models. The applicability and capability of this complex amplitude model is demonstrated with simulations of peritectic solidification and grain growth exhibiting the emergence of secondary phase structures.
\end{abstract}


\maketitle

\section{Introduction}
Understanding complex phenomena during microstructural and phase evolution in materials and condensed matter systems, particularly those associated with system elasticity and plasticity, is at the heart of materials science research. In situ investigation of these phenomena is difficult by experimental means and our theoretical understanding of some of the underlying mechanisms at work is often incomplete, mainly due to the non-equilibrium nature and multiple scales on which these physical mechanisms operate. The design of engineering materials can thus benefit from tractable, yet fundamental, models that capture the full spectrum of microstructural phenomena.

To date, the most successful microstructural modeling approach has come from the use of phenomenologies that have their origins in Ginzburg-Landau and Cahn-Hilliard theories. These models intrinsically operate on the length and time scales relevant to most microstructural processes, i.e., mesoscopic, where information from shorter time and length scales is introduced through effective parameters. The most popular approache is the phase-field (PF) method. This method, notably, has seen great success in the area of solidification~\cite{Karma98,Provatas98,Echebarria04,Greenwood04,Rappaz03,Boettinger99,Granasy03}.

Over the last decade, another class of phase-field models has emerged, i.e., the phase-field-crystal (PFC) model~\cite{Elder02,Elder07}. Unlike its traditional counterpart, the PFC method is an atomic-scale modeling formalism, operating
on atomistic length scales and diffusive time scales. The free energy of PFC models is minimized by periodic fields. As such, the method self-consistently incorporates elasticity, multiple crystal orientations and topological defects. It is rapidly becoming the methodology of choice when investigating atomistic scale effects over diffusive time scales. It has been formally shown, by Elder and coworkers~\cite{Elder07} and Jin and Khachaturyan~\cite{Jin06}, that PFC and PFC-type models, respectively, can be derived from classical density functional theory (CDFT). The PFC method has been successfully applied in the description of solidification~\cite{Tegze09}, spinodal decomposition~\cite{Elder07}, elasto-plasticity~\cite{Stefanovic06}, thin film growth and island formation~\cite{Huang08}, crystal nucleation and polymorphism~\cite{Granasy10,Toth10}, amorphous or glassy states ~\cite{Berry08,Archer12}, among many others.

Most recently, an improved variant of the PFC model has emerged that allows one to control complex crystal structures and their equilibrium coexistence with bulk liquid, i.e., the so-called structural PFC ({\it XPFC}) models. Greenwood {\it et al.}~\cite{Greenwood10,Greenwood11} accomplished this by introducing a class of multi-peaked, two-point direct correlation functions in the free energy functional that contained some of the salient features of CDFT, yet were simplified to be numerically efficient. This XPFC formalism was later extended to binary~\cite{GreenwoodOfori11} and $N$-component~\cite{Ofori-Opoku13} alloying systems, and applied to phenomena such as dendritic and eutectic solidification~\cite{Ofori-Opoku13}, elastic anisotropy ~\cite{GreenwoodOfori11}, solute drag \cite{Greenwood12}, quasi-crystal formation \cite{Rottler12}, solute clustering and precipitation mechanisms in Al-Cu~\cite{Fallah12} and Al-Cu-Mg~\cite{Fallah13,Ofori-Opoku13} alloys, and 3D stacking fault structures in fcc crystals \cite{Berry12}.

Coarse-graining approaches have recently shown that PFC-type models can be used to derive the form of traditional PF models, expressed, however, in the form of complex order-parameters, which makes it possible to simulate different crystal orientations and defect structures on mesoscopic length and time scales. These amplitude models, remarkably, retain many salient atomistic level phenomena, making them prime candidates for multiple scale modeling of microstructure phenomena. Recent amplitude descriptions have been used to describe anisotropic surface energy of crystal-melt interfaces in pure materials and alloys~\cite{Wu07,Majaniemi09,Provatas10}, solidification of multiple crystallites using an adaptive mesh~\cite{Athreya07}, island and quantum dot formation~\cite{Huang08}, segregation and alloy solidification~\cite{Elder10,Spatschek10}, grain boundary premelting~\cite{Spatschek13,Adland13}, and lattice pinning effect on solid-liquid interfaces~\cite{Huang13}. However, these for the most part, have involved pure materials or binary alloys where both elements had the same crystal structure, with calculations based on a single mode approximation of the system free energy functional, i.e., the correlation function was approximated by single peak function.

The purpose of this work is to apply a new coarse-graining approach to the recent XPFC formalism. Recent studies involving coarse-grained PFC models suggest that an amplitude model capable of describing multiple crystal structures and elasto-plastic effects will be valuable in elucidating atomistic scale interactions, material properties and dynamic processes at the mesoscale, as well as motivating more consistent derivations of mesoscale continuum models, such as PF models. Here, we present the amplitude expansion of the XPFC model of the single component system used in \cite{Greenwood10,Greenwood11}, for two-dimensional (2D) square-triangular structures. At the core of our approach is a Fourier method applied to the excess part of the free energy functional, coupled to the volume-averaging technique described in Refs.~\cite{Majaniemi07,Majaniemi09,Provatas10}. After derivation of the corresponding coarse-grained free energy functional, we perform dynamic simulations illustrating solidification and subsequent coarsening, peritectic growth and solid-solid interactions between different crystal structures.

The remainder of this paper is organized as follows. We begin by reviewing the free energy functional of the XPFC model in Sec.~\ref{xpfc-energy}. Section~\ref{amplitude-model} goes through the various steps of generating a complex amplitude free energy functional, including the construction of an appropriate density expansion, then a brief remark on the volume-averaging technique and finally the coarse-graining of the XPFC free energy functional. The dynamics of the set of amplitude equations are discussed in Sec.~\ref{dynamics}, followed by numerical illustrations of the model in Sec.~\ref{amplitude-apps}.

\section{Free energy functional from CDFT}
\label{xpfc-energy}
This section reviews the free energy functional used in Greenwood {\it et al.}~\cite{Greenwood10,Greenwood11}. Particularly, we highlight the excess term in the free energy and examine its correlation kernel, which plays a central role in obtaining different crystal structures in the XPFC model. This is followed by a discussion of the equilibrium properties of the model.

\subsection{XPFC free energy functional of a single component system}
The free energy functional for the XPFC model is derived from the classical density functional theory of Ramakrishnan and Yussouff~\cite{Ramakrishnan79}, containing two contributions. The first is an ideal energy which drives the system to constant homogenous fields, e.g. liquid. The second contribution is an excess term in particle interactions, truncated at the two-particle interaction, which drives the system to be minimized by periodic fields, i.e., solid. In dimensionless form, the resulting XPFC free energy functional can be written as~\cite{Greenwood10,Greenwood11}
\begin{align}
\frac{F}{k_B T\rho_o} &=
\int d\mbfr\bigg\{\frac{F_{id}}{k_B T\rho_o} + \frac{F_{ex}}{k_B T\rho_o} \bigg\}.
\label{structFunc_Pure}
\end{align}
where,
\begin{align}
\frac{F_{id}}{k_B T\rho_o}&=\frac{n^2}{2} - \eta\frac{n^3}{6} + \chi\frac{n^4}{12} \nonumber\\
\frac{F_{ex}}{k_B T\rho_o}&=- \frac{n}{2}~\int d\mbfr^\prime C_2(| \mbfr-\mbfr^\prime|)~n(\mbfr^\prime).
\label{idealExcess}
\end{align}
Here, $n$ is the dimensionless number density, $k_B$ is the Boltzmann constant, $T$ is the temperature and $\rho_o$ is a reference liquid density of the system. $\eta$ and $\chi$ are constants, formally equal to unity, however as discussed in ~\cite{Greenwood11}, deviations from unity allow for better tuning to the full ideal energy, can aid in the mapping to thermodynamic parameters, and can physically be motivated from the contributions of the lowest-order component of higher-order particle correlation functions~\cite{Huang10}. Finally, $C_2(| \mbfr-\mbfr^{\prime}|)$ is the direct two-point correlation function at the reference density $\rho_o$. The construction of this latter expression is what differentiates the XPFC from other PFC variants. We briefly review this next.

\subsection{Correlation function, $C_2(|\mbfr -\mbfr^{\prime}|)$}
The correlation kernel for the XPFC model is constructed in Fourier space, since real space convolutions are simply multiplicative in Fourier space. This also makes the XPFC formalism better equipped for simulations using spectral methods. The correlation function $C_2(|\mbfr -\mbfr^{\prime}|)$ defined at the reference density $\rho_o$, is denoted as $\hat{C}_2(|\mbfk|)$ in Fourier space. A reciprocal space peak of $\hat{C}_2(|\mbfk|)$~\cite{Greenwood11}, for a given mode, $j$, i.e., a peak corresponding to a {\it family} of planes of a desired crystal structure, is denoted by
\beq
\hat{C}_{2j}=e^{-\frac{\sigma^2k_j^2}{\rho_j\beta_j}}e^{-\frac{(k-k_j)^2}{2\alpha^2_j}}.
\label{CorrFunc}
\eeq
The first exponential in Eq.~(\ref{CorrFunc}) sets the temperature scale via a Debye-Waller prefactor that employs an effective temperature parameter, $\sigma$. $\rho_j$ and $\beta_j$ represent the planar and atomic densities, respectively, associated with the family of planes corresponding to mode $j$. These parameters are formally properties of the crystal structure, but can be exploited as constants for convenience and fitting purposes~\cite{GreenwoodOfori11,Ofori-Opoku13}. The second exponential sets the spectral peak position at $k_j$, where $k_j$ is the inverse of the interplanar spacing for the $j^{\rm th}$ family of planes in the unit cell of the crystal structure. Unlike spectral Bragg peaks resulting from diffraction experiments for single crystals, here, each peak is represented by a Gaussian function, with $\alpha_j$ being the width of peak $j$. The $\{\alpha_j\}$ have been shown in Ref.~\cite{Greenwood11} to set the elastic and surface energies and their anisotropic properties. Including only those peaks of the most dominant family of planes, the total correlation function for the crystal structure of interest, $\hat{C}_2$, is then defined by the {\it numerical  envelope} of all peaks $\hat{C}_{2j}$.

Finally, a comment about the $\mbfk = 0$ mode of the correlation function. This mode is the infinite wavelength mode and sets the bulk compressibilities of the system. For simplicity, in Refs.~\cite{Greenwood10,Greenwood11}, the value of the $\mbfk = 0$ was set to zero. A nonzero amplitude at $\mbfk = 0$, however, merely shifts the local free energy at densities away from the reference density ($\rho_o$), thereby causing a compression of the phase diagram about the reference density~\cite{Greenwood11}. This, however, does not alter the stability of the equilibrium crystal structure, since the correlation kernel is constructed about the reference density. It is noteworthy that in addition to setting some bulk properties of the system, the $\mbfk = 0$ will also have an effect on surfaces separating bulk phases, e.g. surface energy. Therefore, in the following, our coarse-graining procedure is done in a general manner that considers a nonzero $\mbfk = 0$, admitting another degree of freedom in mapping to thermodynamic properties of the XPFC model.

\subsection{Equilibrium properties}
The free energy of Eq.~(\ref{structFunc_Pure}) can be shown to yield coexistence of varying crystal structures in equilibrium with liquid~\cite{Greenwood10,Greenwood11}. In 2D, square-liquid and triangle-liquid phases have been studied. In three-dimensions (3D), face-centered cubic (fcc) and liquid, Hexagonal-close packed (hcp) and liquid and body-centered cubic (bcc) and liquid have been demonstrated with single and two-peaked kernels. Furthermore, the free energy of Eq.~(\ref{structFunc_Pure}) can also yield peritectic systems in both 2D and 3D, where multiple solid phases can coexist with liquid. These peritectic systems are comprised of square-triangle-liquid and fcc-bcc-liquid in 2D and 3D, respectively.

\begin{figure}[htb]
\resizebox{3.5in}{!}{\includegraphics{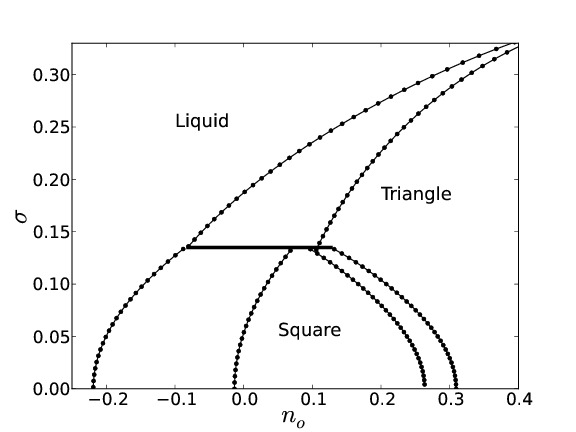}}
\caption{Phase diagram resulting from the minimization of the free energy of Eq.~(\ref{structFunc_Pure}) for 2D structures expanded  in a two-mode approximation. Parameters: $k_{10}=2\pi$ and $k_{11}=2\pi\sqrt{2}$, and $\rho_{10}=1$, $ \beta_{10}=4$, $\alpha_{10}=1$ and $\rho_{11}=1/\sqrt{2}$, $\beta_{11}=4$, $\alpha_{11}=1$. The emergent square and triangle structures have dimensionless lattice spacings of $a_{sq}=1$ and $a_{tri}=2/\sqrt{3}$ respectively.}
\label{fig:PhaseDiagrams}
\end{figure}

Figure~\ref{fig:PhaseDiagrams}, shows a sample phase diagram resulting from minimization of the free energy in Eq.~(\ref{structFunc_Pure}), here for 2D structures. The phase diagram is a result of an input correlation kernel corresponding to a square crystal structure. To stabilize a square crystal structure, the correlation function requires two peaks, $k_{10}$ and $k_{11}$, corresponding to the first two primary family of planes for a square crystal structure. We choose, $k_{10}=2\pi$ and $k_{11}=2\pi\sqrt{2}$ and set, $\rho_{10}=1$, $ \beta_{10}=1$, $\alpha_{10}=1$ and $\rho_{11}=4$, $\beta_{11}=4$, $\alpha_{11}=1$. To stabilize the triangular structure, a single primary peak is sufficient, since additional peaks have a negligible effect on the total energy~\cite{Greenwood11}. We re-scale the position of that peak to be commensurate with the $k_{10}$ peak of the square. In doing so, the two-peaked square correlation function can simultaneously permit square and triangular structures, where the structure with the minimum energy can be parameterized by the average density, $n_o$, and temperature parameter, $\sigma$. After re-scaling, the emergent crystal structures will have dimensionless lattice spacings of $a_{sq}=1$ and $a_{tri}=2/\sqrt{3}$, respectively. To construct a phase diagram, a density mode approximation is introduced for each of the crystal structures of interest, inserted into the free energy and after following standard minimization techniques (see Appendix of Ref.~\cite{GreenwoodOfori11}), the phase diagram shown in Fig.~\ref{fig:PhaseDiagrams} is attained.

Next we shall use the 2D system just discussed to construct a complex order-parameter model via a coarse-graining technique.

\section{Complex order-parameter model: 2D square-triangle structures}
\label{amplitude-model}
Recently, numerous works have been published that perform amplitude expansions, particularly of PFC-type models. The main approaches that have been used are: the multiple scale analysis~\cite{Yeon10,Elder10,Huang10,Huang13}, volume-averaging method~\cite{Majaniemi07,Majaniemi09,Provatas10} and the renomarlization group (RG) approach~\cite{Goldenfeld05,Goldenfeld06,Athreya06,Athreya07}, with the multiple scale method being the most widely applied across disciplines. Older works where these expansions have been performed directly on CDFT models, like the work of Haymet and Oxtoby~\cite{Haymet81,Oxtoby82} and Lakshmi{\it et al.}~\cite{Lakshmi88} fall under the volume-averaging method. Others still, e.g. Kubstrup {\it et al.}~\cite{Kubstrup96}, fall under the multiple scale analysis. The central theme in all these techniques is that the density can be separated into so-called ``fast'' length scales, where the density oscillates rapidly, and ``slow'' length scales, where the amplitudes of the oscillations vary slowly with respect to the rapidly varying oscillation of the density. Beyond this, each method has its own additional underlying assumptions and approximations.

A noteworthy consideration is the validity or accuracy of the various methods in arriving at the same self consistent system of equations. Namely, the multiple scale analysis and RG methods operate on the PFC equations of motion, after which the coarse-grained free energy functional is derived. The volume-averaging method can operate on both the PFC  free energy functional and the dynamical equations, however it has been implemented for the most part at the energy functional level of the PFC or CDFT free energy functionals. A point of criticism against the volume-averaging method, has been the lack of a covariant gradient operator~\cite{Gunaratne94} in the amplitude equations. However, this problem can be circumvented by expanding to higher order in the amplitude expansion. In previous implementations, only a second order expansion was taken of the ``slow'' variables (i.e., the amplitudes)~\cite{Majaniemi09,Provatas10}. While to second order, surface energy calculations can be performed quite quantitatively, dynamic simulations become fixed to certain orientations. It has been shown~\cite{Ofori-OpokuUP}, that an expansion of amplitudes appearing in the excess term to at least fourth order is necessary in order to recover the lowest order covariant gradient operator in the volume-averaging approach.

This work will use the volume-averaging technique to perform calculations, in conjunction with a novel method to handle the excess term in the PFC free energy functional. To begin, we first discuss the separation of scales via an expansion of the density that describes two crystal lattices. After discussing the density expansion, we briefly outline the basic features of the volume-averaging method. As the method has been published elsewhere, the outline given will highlight the important concepts of the method, after which it is applied to the ideal portion of the free energy. Finally we introduce the method of handling the excess term, which completes the amplitude derivation for the 2D XPFC model.

\subsection{Density expansion in two lattices}
\label{dens-expan}
The PFC suite of models for a pure material contain only a single dimensionless density field, $n$. A self-consistent method of putting forth a density expansion which incorporates multiple crystal structures is nontrivial. For the XPFC, in 2D these crystal structures are crystals of triangular and square symmetry. Kubstrup {\it et al.}~\cite{Kubstrup96} in a study of pinning effects between fronts of hexagonal (i.e., triangular) and square phases, have proposed a construction through which variable phases can be described by a single expansion definition. This density expansion, for the XPFC model, can be written as,
\beq
n(\mbfr) = n_o(\mbfr) + \sum_j^{6} A_j(\mbfr)e^{i\mbfk_j\cdot\mbfr} + \sum_m^{6} B_m(\mbfr)e^{i\mbfq_m\cdot\mbfr} + c.c.,
\label{densExpan12}
\eeq
where $n_o(\mbfr)$ is the dimensionless average density and is a ``slow'' variable, ``$c.c.$'' denotes the complex conjugate, $\{A_j\}$ represent the amplitudes describing the first mode of our structures, while all $\{B_m\}$  represent the amplitudes for the second mode. Like the dimensionless average density, the amplitudes are also ``slow'' variables.

\begin{figure}[htb]
\centering
	\resizebox{3.5in}{!}{\includegraphics{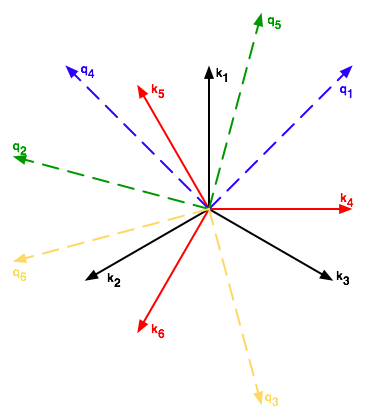}}
    \caption{(color online) Schematic representation of, mis-oriented by 30 degrees, the vector sets $\{\mbfk_{1},\mbfk_{2},\mbfk_{3}\}$ and $\{\mbfk_{4},\mbfk_{5},\mbfk_{6}\}$, respectively, which form a resonant set and compromise two triangular lattices. Vectors $\mbfk_{i}$ and $\mbfk_{i+3}$ ($i=1,2,3$) are orthogonal to each other, forming the first mode of the square correlation kernel. The other set of vectors, $\{\mbfq_{m}\}$, correspond to different orientations of the second mode of the correlation kernel necessary to stabilize the square structure in the XPFC formalism, and are formed from a linear combination of the orthogonal pairs from the two triangular sets.}
\label{fig:RecipVectors12}
\end{figure}

Note that only the first mode $\{\mbfk_j\}$ was considered by Kubstrup {\it et al.}~\cite{Kubstrup96} in the study of pattern formation, while in Eq. (\ref{densExpan12}) we have included both the zeroth mode $n_{o}$ (as a result of density conservation) and also the second mode $\{\mbfq_m\}$, which is needed for stabilizing the square structure in the XPFC formalism. The density expansion we construct can be schematically inferred from Fig.~\ref{fig:RecipVectors12} in terms of the required set of reciprocal lattice vectors. Figure~\ref{fig:RecipVectors12} represents the reciprocal lattice vectors that enter the density expansion in Eq.~(\ref{densExpan12}), having two interlaced triangular structures mis-oriented by $30^\circ$, i.e., vectors $\mbfk_1,\mbfk_2,\mbfk_3$ and $\mbfk_4,\mbfk_5,\mbfk_6$ each forming a triangular lattice, respectively. It will be useful in what follows that the property of resonance is satisfied by these two vector sets. Resonance between density waves is satisfied when $\mbfk_1+\mbfk_2+\mbfk_3=0$ and $\mbfk_4+\mbfk_5+\mbfk_6=0$.  The square structure can be partly constructed from combinations of the reciprocal lattice vectors of the two triangular sets. For example, $\mbfk_1$ and $\mbfk_4$ (which are orthogonal, i.e., $\mbfk_1\cdot\mbfk_4=0$) represent the first mode of a square lattice, while the second mode of the square can be constructed from a linear combination, such as $\mbfq_1=\mbfk_1+\mbfk_4$ and $\mbfq_4=\mbfk_1-\mbfk_4$. Analogous associations can be made for the second and third set of square lattices which arise from the two interlaced triangular lattices. In total, the density expansion for a system described by the vectors of Fig.~\ref{fig:RecipVectors12} amount to 12 vectors and therefore 12 complex amplitudes.
\begin{figure}[htb]
\centering
	\resizebox{3.5in}{!}{\includegraphics{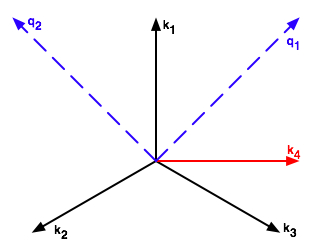}}
    \caption{(color online) Second schematic representation of the reciprocal set of basis vectors which comprise a density simultaneously describing crystals with square and triangular symmetry respectively. Vectors $\mbfk_{1},\mbfk_{2},\mbfk_{3}$ form a resonant set and compromise a single triangular lattice. Vectors $\mbfk_{1}$ and $\mbfk_{4}$ are orthogonal forming the first mode of the square correlation kernel. The other set of vectors, \{$\mbfq_{m}$\} dashed-dotted, comprise the second mode of the square correlation kernel.}
\label{fig:RecipVectors6}
\end{figure}

The expansion described by Eq.~(\ref{densExpan12}) and the vectors of Fig.~\ref{fig:RecipVectors12}, each corresponding to one of the 12 amplitudes may prove to be intractable or at the least tedious and cumbersome to deal with. A simpler more intuitive expansion, is also proposed here as a comparison. This is illustrated by the reciprocal lattice vectors of Fig.~\ref{fig:RecipVectors6}. Unlike the previous expansion, this expansion requires 6 vectors and hence 6 amplitudes. At first glance, there seems to be a limited number of degrees of freedom afforded to us by an expansion of this kind, in particular the pre-set orientation of the square structure that is constrained to the $\{\mbfk_1,\mbfk_4\}$ direction. This and other nuances that may exist between the two expansions may be ascertained through numerical simulations. For convenience, we will be using this latter expansion in our derivation to follow. In Appendix~\ref{energy12vector}, we also report the complex amplitude model derived from the 12 amplitude expansion described by Eq.~(\ref{densExpan12}). The simpler density expansion based on the vectors of Fig.~\ref{fig:RecipVectors6} is written as
\beq
n(\mbfr) = n_o(\mbfr) + \sum_j^{4} A_j(\mbfr)e^{i\mbfk_j\cdot\mbfr} + \sum_m^{2}B_m(\mbfr)e^{i\mbfq_m\cdot\mbfr} + c.c.
\label{densExpan6}
\eeq

\subsection{Volume-averaging technique for coarse-graining}
\label{volume-averaging}
As mentioned in the previous section, the amplitudes $\{A_j\}$ and $\{B_m\}$ along with the dimensionless average density $n_o$, are all slowly varying on atomic scales. After inserting the density expansion of Eq.~(\ref{densExpan6}) into the XPFC free energy terms of Eq.~(\ref{idealExcess}), to lowest order the terms that will survive in the coarse-graining procedure are those where the oscillating exponential phase factors vanish. In particular, under coarse-graining, the free energy effectively becomes a series of terms with ``slow'' variables multiplying phase factors of the form $e^{i\Delta Q_{l}\cdot\mbfr}$, where $\Delta Q_{l}$ are sums or differences in the reciprocal lattice vectors. As in all coarse-graining approaches, the lowest order approximation, i.e., so-called ``quick and dirty'' approach~\cite{Goldenfeld05}, amounts to the situation where the only surviving coarse-grained terms result from all $\Delta Q_{l}\equiv 0$, i.e., where a resonant condition is satisfied. This is the standard condition from the symmetry requirement of translational invariance of the total free energy~\cite{Shih87}.

Formally, coarse-graining can be done by the volume-averaging method using a convolution operator~\cite{Majaniemi09}, which can be defined by,
\begin{align}
\label{volume-avg-1}
\langle  f(\mbfr) \rangle_V \equiv \int_{-\infty}^{\infty} d\mbfr^{\prime} f(\mbfr^{\prime}) \xi_V(\mbfr-\mbfr^{\prime}),
\end{align}
where $f(\mbfr^{\prime})$ is the function being coarse-grained, for our purposes collections of ``slow'' variables or ``slow'' variables multiplied by phase factors and $V$ is the coarse-graining volume, i.e., typically the volume of a unit cell. The function $\xi_V$ in the integrand of Eq.~(\ref{volume-avg-1}) is a smoothing function that is normalized to unity, i.e.,
\begin{align}
\label{volume-avg-2}
\int_{-\infty}^{\infty} d\mbfr\,\xi_{V}(\mbfr-\mbfr^{\prime})\equiv 1.
\end{align}
In the long wavelength limit, $L_{slow} \gg L \gg a$ where $L~\sim V^{1/d}$, in $d$-dimensions, where $L_{slow}$ is the length scale of variation of the ``slow'' variables (i.e. microstructural features), and $a$ is the equilibrium lattice spacing. This condition implies that the function $\xi_V(\mbfr)$ varies on dimensions much larger than the lattice constant, e.g. $a = 2\pi/|\mbfk_j|$, but much less than the length scale of variation of the average density and amplitudes. Equation~(\ref{volume-avg-1}) is formally applied by changing the dependent variable in the free energy functional from $\mbfr$ to $\mbfr^{\prime}$, multiplying the resulting free energy by the left hand side of Eq.~(\ref{volume-avg-2}) (i.e., 1) and inverting the order of integration, thus arriving at a series of terms of the form of Eq.~(\ref{volume-avg-1}). Equation~(\ref{volume-avg-1}) defines a noninvertible limiting procedure that can be used to average a function over some volume. The reader is referred to Refs.~\cite{Majaniemi07,Majaniemi09,Ofori-Opoku13} for greater detail about the application of the volume-averaging convolution operator.

\subsection{Coarse-graining the ideal term}
\label{ideal-coarse}
Inserting the density expansion of Eq.~(\ref{densExpan6}) into the ideal portion, $F_{id}$, of the free energy, and coarse-graining as described in the preceding section, yields to lowest order in the average density and amplitudes,
\begin{widetext}
\begin{align}
\frac{F_{id}^{cg}}{k_B T\rho_{o} V} &= \int d\mbfr\, \Bigg\{ \frac{n_o^2}{2} - \eta\frac{n_o^3}{6} + \chi\frac{n_o^4}{12} + \lt(1-\eta\,n_o+ \chi\,n^2_o\rt)\,\bigg(\sum_{j}^4|A_j|^2 + \sum_{m}^2|B_m|^2\bigg) \nline
&-(\eta-2\chi n_o)\lt[A_1A_2A_3 + A_1^*A_4^*B_1 + A_1A_4^*B_2^* + c.c. \rt]\nonumber\\
&+\frac{\chi}{2}\lt[\sum_{j}^4A_j^2(A^*_j)^2 +  \sum_{m}^2B_m^2(B^*_m)^2\rt]
+ 2\chi\lt[\lt(\sum_j^4\sum_{m>j}^4|A_j|^2|A_m|^2 + \sum_j^4\sum_m^2|A_j|^2|B_m|^2 + |B_1|^2|B_2|^2\rt)\rt] \nline
&+\chi\lt[2A_2^*A_3^*A_4^*B_2^* + 2A_2^*A_3^*A_4B_1^* + A_1^2B_1^*B_2^* + A_4^2B_1^*B_2 + c.c. \rt] \Bigg\},
\label{ampFid}
\end{align}
\end{widetext}
where $^*$ denotes the complex conjugate and in the coarse-grained free energy the spatial variable, $\mbfr$, is scaled by the lattice constant $a$. As alluded to earlier, the correlation-containing excess term has received some attention in the coarse-graining of PFC models. In the following section, we introduce a general Fourier method to coarse-grain this term in the context of the present XPFC model.

\subsection{Coarse-graining the excess term}
\label{excess-correlation}
We first rewrite the correlation kernel in its Fourier series representation, i.e.,
\beq
C_2(|\mbfr-\mbfr^{\prime}|)=\int d\mbfk \, \hat{C}_2(|\mbfk|)\,e^{i\mbfk\cdot\mbfr}\,e^{-i\mbfk\cdot\mbfr^{\prime}}.
\eeq
The convolution term, the integral over $d\mbfr^{\prime}$, in the free energy involving the excess term then becomes,
\begin{align}
\label{excessConv-1}
{\cal G} &=\int d\mbfr^{\prime} C_2(|\mbfr-\mbfr^{\prime}|) n(\mbfr^{\prime}) \nonumber \\
&=\int d\mbfr^{\prime}\,\int d\mbfk\,\hat{C}_2(|\mbfk|)e^{\mbfi\mbfk\cdot\mbfr}\,e^{-\mbfi\mbfk\cdot\mbfr^{\prime}}\,
n(\mbfr^{\prime}).
\end{align}
Next we take the Taylor series expansion of the correlation function around $\mbfk=0$, i.e., the infinite wavelength mode of the correlation, to all orders\footnote{It is tacitly assumed in doing so that the expansion, Eq.~(\ref{corr-expan-k}), is convergent over a large enough window of $\mbfk$-values.}. This expansion can be compactly written as,
\beq
\hat{C}_2(|\mbfk|) = \sum_{l=0}^{\infty}\frac{1}{l!}(\mbfk)^{l}\frac{\partial^{l}\hat{C}_2}{\partial \mbfk^{l}}\bigg|_{\mbfk=0}.
\label{corr-expan-k}
\eeq
This functional Taylor series expansion is formally exact, as it goes to all orders. Substituting the density expansion, Eq.~(\ref{densExpan6}), into the convolution term of the free energy and employing the definition of the Fourier transform yields
\begin{align}
\label{excessConv-3}
{\cal G} &=\int
d\mbfk\,\sum_{l=0}^{\infty}\vartheta_{l}~(\mbfk)^{l}\hat{n}_o(\mbfk)e^{\mbfi\mbfk\cdot\mbfr}\\
&+\int d\mbfk\,\sum_{l=0}^{\infty}\vartheta_{l}~(\mbfk)^{l}
\sum_j^{4} \hat{A}_j(\mbfk-\mbfk_j)e^{\mbfi\mbfk\cdot\mbfr}\nonumber \\
&+\int d\mbfk\,\sum_{l=0}^{\infty}\vartheta_{l}~(\mbfk)^{l}
\sum_m^{2}\hat{B}_m(\mbfk-\mbfq_m)e^{\mbfi\mbfk\cdot\mbfr} + c.c.\nonumber,
\end{align}
where we have made the following definition,
\begin{align}
\vartheta_l &= \frac{1}{l!}\frac{\partial^{l}\hat{C}_2}{\partial \mbfk^{l}}\bigg|_{\mbfk=0},
\label{corr-params}
\end{align}
and $\hat{n}_o$ , $\hat{A}_j$ and $\hat{B}_m$ are the corresponding Fourier components of the average density and amplitudes respectively. Next we re-sum the correlation function for the average density part of the convolution term in Eq.~(\ref{excessConv-3}), and make consecutive changes of variables, i.e., $\mbfk^{\prime}=\mbfk-\mbfk_{j}$ and then $\mbfk^{\prime}=\mbfk-\mbfq_{m}$, for the second and third terms of Eq.~(\ref{excessConv-3}), respectively. Following these steps, we arrive at
\begin{align}
\label{excessConv-4}
&{\cal G} = \lt[\hat{C}_2(|\mbfk|)\hat{n}_{o}(\mbfk)\rt]_{\mbfr}\\
&+\sum_j^{4}\int\,d\mbfk^{\prime}\sum_{l=0}^{\infty}\vartheta_{l}~(\mbfk^{\prime}+\mbfk_{j})^{l}\,
\hat{A}_j(\mbfk^{\prime})e^{\mbfi\mbfk^{\prime}\cdot\mbfr}e^{\mbfi\mbfk_j\cdot\mbfr} \nonumber \\
&+\sum_{m}^{2}\int\,d\mbfk^{\prime}\sum_{l=0}^{\infty}\vartheta_{l}~(\mbfk^{\prime}+\mbfq_{m})^{l}
\hat{B}_m(\mbfk^{\prime})e^{\mbfi\mbfk^{\prime}\cdot\mbfr}e^{\mbfi\mbfq_m\cdot\mbfr} \nonumber \\
&+ c.c.\nonumber
\end{align}
Applying the definition of the Fourier transform to  the second and third terms on the RHS of Eq.~(\ref{excessConv-4}) yields,
\begin{align}
\label{excessConv-5}
{\cal G} &=\lt[\hat{C}_2(|\mbfk|)\hat{n}_{o}(\mbfk)\rt]_{\mbfr}
+\sum_{j}^{4}e^{\mbfi\mbfk_j\cdot\mbfr}\lt[\hat{C}_2(|\mbfk+\mbfk_j|)\hat{A}_{j}(\mbfk)\rt]_{\mbfr}\nonumber \\
&+\sum_{m}^{2}e^{\mbfi\mbfq_m\cdot\mbfr}\lt[\hat{C}_2(|\mbfk+\mbfq_m|)\hat{B}_{m}(\mbfk)\rt]_{\mbfr}
+c.c.,
\end{align}
where $\lt[~\rt]_{\mbfr}$ denotes the inverse Fourier transform. Equation~(\ref{excessConv-5}) represents the total convolution term of the excess free energy.

To complete the coarse-graining of the excess term, we multiply the convolution term in Eq.~(\ref{excessConv-5}) by the expansion of the density field, i.e., $n\,\mathcal{G}$, and apply the convolution operator in Eq.~(\ref{volume-avg-1}), to obtain the lowest order result;
\begin{align}
\label{ampFex}
\frac{F_{ex}^{cg}}{k_B T\rho_{o} V} &=\int d\mbfr\, \Bigg\{
-\frac{n_o}{2}\lt[\hat{\xi}_{V}(\mbfk)\hat{C}_2(|\mbfk|)\hat{n}_{o}(\mbfk)\rt]_{\mbfr}\\
&-\frac{1}{2}\sum_{j}^{4}A_{j}^{*}\lt[\hat{C}_2(|\mbfk+\mbfk_j|)\hat{A}_{j}(\mbfk)\rt]_{\mbfr}\nonumber \\
&-\frac{1}{2}\sum_{m}^{2}B_{m}^{*}\lt[\hat{C}_2(|\mbfk+\mbfq_m|)\hat{B}_{m}(\mbfk)\rt]_{\mbfr} +c.c.\Bigg\},\nonumber
\end{align}
where $\hat{\xi}_V$ is the convolution operator in Fourier space, which filters out $\hat{C}_2$ oscillations beyond its $\mbfk=0$ peak with some decay range in Fourier space. The explicit derivation of this term is discussed in Sec.~\ref{average-density}.

Several things are worth noting in Eq.~(\ref{ampFex}). It becomes evident that the rotational invariance nature of a system, afforded through the covariant gradient operator in real space, is manifested here in the correlation kernel, which has as input a shifted wavenumber for the respective modes being considered. This shifted wavenumber samples low-$\mbfk$ value deviations (long wavelength limit) around the peaks of the original correlation function. This essentially treats each reciprocal space peak of the original correlation kernel as a corresponding effective ``$\mbfk=0$'' mode. Like the microscopic XPFC model, the full correlation kernel, in this amplitude formalism, is the {\it numerical envelope} of all reciprocal space peaks included to represent the crystal structural of interest.

Combining Eq.~(\ref{ampFid}) and (\ref{ampFex}), we arrive at a complete coarse-grained free energy for the structural PFC model of the form,
\begin{widetext}
\begin{align}
F^{cg} &= \int d\mbfr \Bigg\{\frac{n_o^2}{2} - \eta\frac{n_o^3}{6} + \chi\frac{n_o^4}{12} + \lt(1-\eta\,n_o+ \chi\,n^2_o\rt)\,\bigg(\sum_{j}^4|A_j|^2 + \sum_{m}^2|B_m|^2\bigg) \nline
&-(\eta-2\chi n_o)\lt[A_1A_2A_3 + A_1^*A_4^*B_1 + A_1A_4^*B_2^* + c.c. \rt]\nonumber\\
&+\frac{\chi}{2}\lt[\sum_{j}^4A_j^2(A^*_j)^2 +  \sum_{m}^2B_m^2(B^*_m)^2\rt]
+ 2\chi\lt[\lt(\sum_j^4\sum_{m>j}^4|A_j|^2|A_m|^2 + \sum_j^4\sum_m^2|A_j|^2|B_m|^2 + |B_1|^2|B_2|^2\rt)\rt] \nline
&+\chi\lt[2A_2^*A_3^*A_4^*B_2^* + 2A_2^*A_3^*A_4B_1^* + A_1^2B_1^*B_2^* + A_4^2B_1^*B_2 + c.c. \rt]\nonumber \\
&-\frac{n_o}{2}\lt[\hat{\xi}_{V}(\mbfk)\hat{C}_2(|\mbfk|)\hat{n}_{o}(\mbfk)\rt]_{\mbfr}-\frac{1}{2}\sum_{j}^{4}A_{j}^{*}\lt[\hat{C}_2(|\mbfk+\mbfk_j|)\hat{A}_{j}(\mbfk)\rt]_{\mbfr} -\frac{1}{2}\sum_{j}^{4} A_{j}\lt[\hat{C}_2(|\mbfk-\mbfk_j|)\hat{A}_{j}(-\mbfk)\rt]_{\mbfr} \nonumber \\
&-\frac{1}{2}\sum_{m}^{2}B_{m}^{*}\lt[\hat{C}_2(|\mbfk+\mbfq_m|)\hat{B}_{m}(\mbfk)\rt]_{\mbfr} -\frac{1}{2}\sum_{m}^{2} B_{m}\lt[\hat{C}_2(|\mbfk-\mbfq_m|)\hat{B}_{m}(-\mbfk)\rt]_{\mbfr}\Bigg\}.
\label{ampEnergy}
\end{align}
\end{widetext}

\subsection{Recovering the amplitude representations of other PFC models}
\label{excess-correlation-CovOp}
Our Fourier method from the above section can also be used to recover the covariant gradient operators found in amplitude expansions of other PFC  models. Here we consider an expansion of the correlation around $\mbfk=0$, in powers of $\mbfk^2$, similar to the standard PFC model of Elder and co-workers~\cite{Elder02,Elder04} but generalized to all orders. This can be compactly written as,
\beq
\hat{C}_2(|\mbfk|) = \sum_{l=0}^{\infty}\frac{1}{l!}(\mbfk^2)^{l}\frac{\partial^{l}\hat{C}_2}{\partial (\mbfk^2)^{l}}\bigg|_{\mbfk=0},
\eeq
We note that this expansion can be generally valid provided the correlation is some well-behaved function and expressible to reasonable accuracy in even powers of $\mbfk$. This is true for most correlations derived from experiments or first principle calculations or those that can be fit to such techniques. An appropriate example is the eighth order fitting of Jaatinen and Ala-Nissila~\cite{Jaatinen10}, which was found to be an accurate and efficient approximation to CDFT. Applying the same arguments leading up to Eq.~(\ref{excessConv-3}) gives
\begin{align}
\label{excessConv-CovOp}
{\cal G} &=\int d\mbfk\,\sum_{l=0}^{\infty}\varepsilon_{l}~(\mbfk^2)^{l}\hat{n}_o(\mbfk)e^{\mbfi\mbfk\cdot\mbfr}\\
&+\int d\mbfk\,\sum_{l=0}^{\infty}\varepsilon_{l}~(\mbfk^2)^{l}\
\sum_j^{4} \hat{A}_j(\mbfk-\mbfk_j)e^{\mbfi\mbfk\cdot\mbfr} \nonumber \\
&+\int d\mbfk\,\sum_{l=0}^{\infty}\varepsilon_{l}~(\mbfk^2)^{l}\sum_m^{2}\hat{B}_m(\mbfk-\mbfq_m)e^{\mbfi\mbfk\cdot\mbfr} + c.c.,\nonumber
\end{align}
where we have made the following definition,
\begin{align}
\varepsilon_l &= \frac{1}{l!}\frac{\partial^{l}\hat{C}_2}{\partial (\mbfk^2)^{l}}\bigg|_{\mbfk=0}.
\end{align}
Using the definition of the Fourier transform on the RHS of Eq.~(\ref{excessConv-CovOp}) leads to
\begin{align}
\label{excessConv-CovOp-2}
{\cal G} &= \sum_{l=0}^{\infty}\varepsilon_{l}\,(-\nabla^2)^{l}\,n_{o}(\mbfr) + \sum_{l=0}^{\infty}\varepsilon_{l}\,(-\nabla^2)^{l}\,
\sum_j^{4} A_j(\mbfr)e^{\mbfi\mbfk_j\cdot\mbfr}\nonumber \\
&+\sum_{l=0}^{\infty}\varepsilon_{l}\,(-\nabla^2)^{l}\,
\sum_m^{2}B_m(\mbfr)e^{\mbfi\mbfq_m\cdot\mbfr} + c.c.
\end{align}
Noting that, $\nabla^2\rightarrow\nabla^2 + 2\mbfi\mbfk_j\cdot\nabla -\mbfk_j^{2}$ (the covariant gradient operator), when Laplacian operators act on terms of the form $A_j(\mbfr)e^{\mbfi\mbfk_j\cdot\mbfr}$, Eq.~(\ref{excessConv-CovOp-2}) becomes
\begin{align}
\label{excessConv-CovOp-3}
{\cal G} &=\sum_{l=0}^{\infty}\varepsilon_{l}(-\nabla^2)^{l}\,n_{o}(\mbfr)\\
&+\sum_{j}^{4}e^{\mbfi\mbfk_j\cdot\mbfr}\sum_{l=0}^{\infty}\varepsilon_{l}\bigg\{-\lt(\nabla^2 + 2\mbfi\mbfk_{j}\cdot\nabla -\mbfk_j^{2}\rt)\bigg\}^{l}\,
 A_j(\mbfr)\nonumber\\
&+\sum_{m}^{2}e^{\mbfi\mbfq_m\cdot\mbfr}\sum_{l=0}^{\infty}\varepsilon_{l}\bigg\{-\lt(\nabla^2 + 2\mbfi\mbfq_m\cdot\nabla -\mbfq_m^{2}\rt)\bigg\}^{l}\,
B_m(\mbfr) \nonumber \\
&+ c.c. \nonumber
\end{align}
Equations~(\ref{excessConv-CovOp-2}) and (\ref{excessConv-CovOp-3}) show that an infinite set of covariant gradient operators (in the long wavelength limit) is needed to accurately capture the salient features, in real space, of a correlation kernel constructed in Fourier space, reflecting that the latter would require an infinite series of square gradient terms to be represented in a traditional PFC form. If we neglect all second mode contributions and truncate the series at $l=2$ in Eq.~(\ref{excessConv-CovOp-3}), we recover the amplitude expansion of the standard PFC model~\cite{Elder10}, after the usual application of the coarse-graining operation. To make contact with the generalized formalism of the previous section, the amplitude terms are rewritten in terms of an inverse Fourier transform via a change of variable, and in Fourier space the resulting correlation kernel expansion is re-summed, resulting in the same coarse-grained free energy form as Eq.~(\ref{ampFex}).

\section{Periodic instability arising from the average density}
\label{average-density}
It turns out that the use of the ``quick and dirty''  or multiple scale method to coarse-grain the standard PFC model leads to a term of the form $n_o (1+\nabla^2)^2 n_o$~\cite{Yeon10,Huang10}, which can become unstable to periodic oscillations in the average density that replicates those of the original density field, $n$, particularly around sharp solid-liquid interfaces. Often a second long wavelength approximation is made to suppress the associated terms responsible for the instability~\cite{Yeon10}. In our approach this instability is self-consistently eliminated through the convolution operator. We qualify this statement by showing the explicit steps required to coarse-grain the average density term in Eq.~(\ref{excessConv-5}).

We start with the form of the correlation contribution of the average density term prior to introducing the volume-averaging kernel of Eq.~(\ref{volume-avg-2}). From Eq.~(\ref{excessConv-5}), we have
\begin{align}
\label{avg-dens-1}
{\cal H} &= -\frac{1}{2}\int\,d\mbfr^{\prime} \,n_{o}(\mbfr^{\prime})\lt[\hat{C}_2(|\mbfk|)\hat{n}_{o}(\mbfk)\rt]_{\mbfr^{\prime}}.
\end{align}
After inserting the volume-averaging kernel, we have
\begin{align}
\label{avg-dens-2}
{\cal H}^{cg} &= -\frac{1}{2}\int\,d\mbfr^{\prime} \int\,d\mbfr\,\lt[\int\,d\mbfq\,\hat{\xi}_{V}(\mbfq)\,e^{\mbfi\mbfq\cdot\mbfr}e^{-\mbfi\mbfq\cdot\mbfr^{\prime}}\rt] \nonumber \\
&\times n_{o}(\mbfr^{\prime})\int\,d\mbfk\,\hat{C}_2(|\mbfk|)\,\hat{n}_{o}(\mbfk)e^{\mbfi\mbfk\cdot\mbfr^{\prime}}.
\end{align}
Here, $\hat{\xi}_{V}$ is the averaging (or convolution) kernel in Fourier space, which restricts the wavenumber $\mbfq$ to small values, i.e., $\mbfq < 1/L$, approximately the same as the first Brillouin zone or similarly the first peak of the correlation function. Note that the average density variable, $n_{o}(\mbfr^{\prime})$, is slowing varying, while the multiplication of $\hat{C}_2(|\mbfk|)\hat{n}_{o}(\mbfk)$ can lead to rapid oscillations due to instabilities caused by the correlation kernel. On scales of the rapidly oscillating term, it is reasonable to take an expansion of $n_{o}(\mbfr^{\prime})$, i.e., $n_{o}(\mbfr^{\prime})\rightarrow n_{o}(\mbfr)$, which allows us to remove it from the integral over $\mbfr^{\prime}$. Next, the noninvertible procedure, described above, occurs by switching the order of integration $d\mbfr$ with $d\mbfr^{\prime}$, after which we integrate the equation with respect to $\mbfr^{\prime}$ yielding,
\begin{align}
\label{avg-dens-4}
{\cal H}^{cg} &= -\frac{1}{2}\int\,d\mbfr\,n_{o}(\mbfr) \int\,d\mbfq\,\hat{\xi}_{V}(\mbfq)\hat{C}_2(|\mbfq|)\hat{n}_{o}(\mbfq)e^{\mbfi\mbfq\cdot\mbfr}\nonumber \\
&= -\frac{1}{2}\int\,d\mbfr\,n_{o}(\mbfr)\lt[\hat{\xi}_{V}(\mbfk)\hat{C}_2(|\mbfk|)\hat{n}_{o}(\mbfk)\rt]_{\mbfr},
\end{align}
where $\lt[~\rt]_{\mbfr}$ denotes the inverse Fourier transform and in the last line we have changed the wavenumber variable for convenience, i.e., $\mbfq\rightarrow\mbfk$. Equation~(\ref{avg-dens-4}), clearly demonstrates that all the small wavelength modes in the original correlation function associated with the periodic instability of $n_o$ are suppressed by convolving with the volume-averaging kernel. In other words, considering the volume-averaging kernel as a filter (in this case a low-pass filter), it smooths or eliminates all the high-mode (small wavelength) peaks resulting from the correlation function. This effectively allows the system to only sample the long wavelength information of the correlation function around $\mbfk=0$.

Equivalently, this can also be motivated from the multiple scale method of coarse-graining. In that method, a small parameter, $\epsilon$~\cite{Huang08,Huang10}, is introduced in a perturbation type expansion which results in the wavenumber being described by $\mbfk \rightarrow \mbfk+\epsilon\mbfK$, where $\mbfK$ represents the large wavelength modes. Considering the long wavelength behavior of the average density, this results in the correlation function being evaluated at $\epsilon\mbfK$, i.e., $\hat{C}_{2}(|\epsilon\mbfK|)$, effectively shifting the modes sampled by the correlation to only those around $\mbfk=0$. It is worth noting that if one simply applies the so-called ``quick and dirty'' approach, of any of the coarse-graining methods when considering a density jump, the average density term will not be coarse-grained, resulting in a term which still possesses the
small scale feature of the original free energy functional.

\section{Dynamics}
\label{dynamics}
Dynamics of the complex order-parameters comprising the coarse-grained free energy follow the usual variational principle of traditional phase-field models. Particularly, the average density, $n_o$, obeys conserved dissipative dynamics, while the amplitudes $A_j$ and $B_m$ follow nonconserved dissipative dynamics
\footnote{While in this derivation the dynamics of $n_o$ follow conserved and diffusive dynamics, in a more complete treatment, the dynamics of $n_o$ can also be treated as a slow spatial but rapidly changing variable through the application of inertial dynamics~\cite{Stefanovic06} to the $n_o$ equation.}.
Specifically we have
\begin{align}
\label{dynamics-no}
\frac{\partial n_o}{\partial t} \!\! &= \!\nabla \! \cdot \! \left( M_{n_o}\nabla \frac{\delta F^{cg}}{\delta\,n_o} \right) + \nabla\cdot\zeta_{n_o},
\end{align}
\begin{align}
\label{dynamics-ampA}
\frac{\partial A_{j}}{\partial t} \!\! &= -M_{A_{j}}\frac{\delta F^{cg}}{\delta\,A_{j}^{*}} + \zeta_{A_j},
\end{align}
and
\begin{align}
\label{dynamics-ampB}
\frac{\partial B_{m}}{\partial t} \!\! &= -M_{B_{m}}\frac{\delta F^{cg}}{\delta\,B_{m}^{*}} + \zeta_{B_m}.
\end{align}
The reader is referred to Appendix~\ref{dynamicsFull} for the full set of explicitly written dynamic equations. The coefficients $M_{n_o}$, $M_{A_{j}}$ and $M_{B_{m}}$ denote the mobility  parameters of the average density and each corresponding amplitude, respectively, and strictly speaking can be functions of the various fields in the free energy functional. We have appended to these equations of motion, the stochastic variables $\zeta_{n_o}$, $\zeta_{A_{j}}$ and $\zeta_{B_{m}}$, which model coarse grained thermal fluctuations acting on the average density and amplitudes, respectively. Formally, they satisfy the fluctuation-dissipation theorem, i.e., $\langle \zeta_{\nu} (\mbfr,t)\rangle=0$ and $\langle \zeta_{\nu} (\mbfr,t) \zeta_{\nu} (\mbfr^{\prime},t^{\prime})\rangle = \Gamma_{\nu} \delta(\mbfr-\mbfr^{\prime}) \delta (t-t^{\prime}) $, where $\nu$ denotes the average density or one of the amplitude fields, with $\Gamma_{\nu} \propto M_{\nu} k_B T $. Huang {\it et al.}~\cite{Huang10} have formally shown how these coarse-grained stochastic variables are derived, in an amplitude equation formalism from dynamic density functional theory through multiple scale analysis. Next we showcase the dynamics properties of the derived XPFC amplitude model through three types of microstructure simulations.
\begin{figure*}[htb]
    \centering
    \begin{tabular}{cccc}
    \includegraphics[width=1.3in]{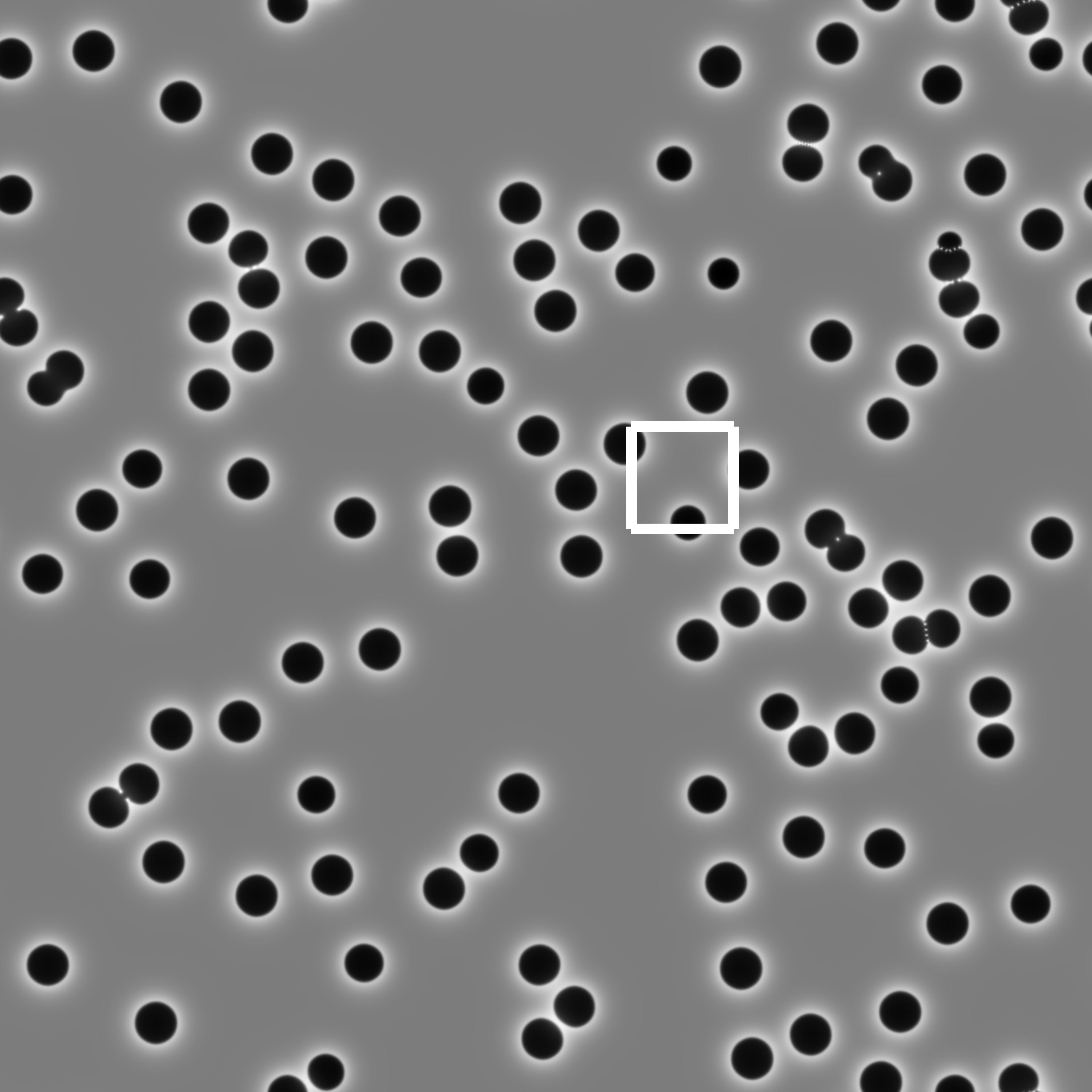}&\includegraphics[width=1.3in]{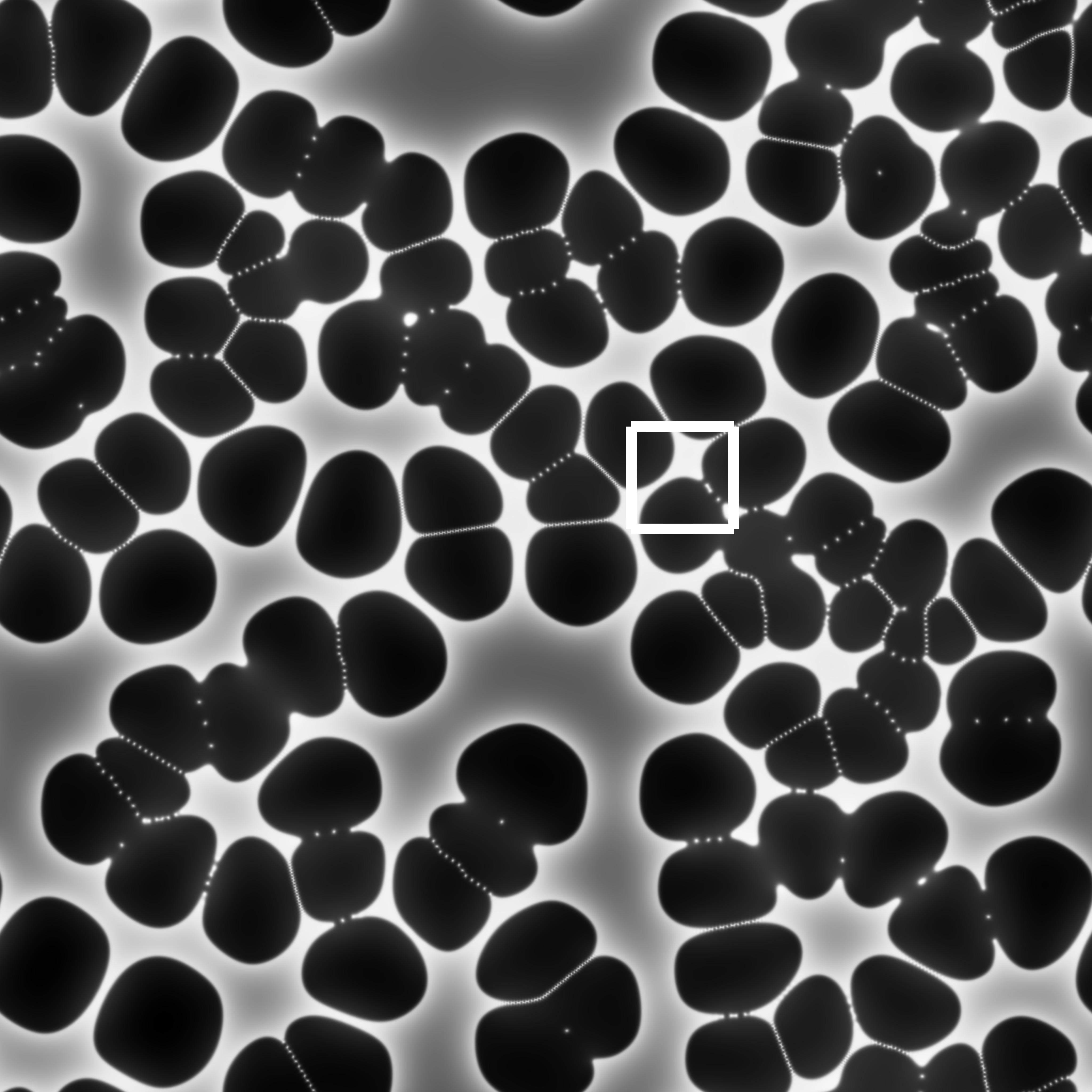}
    &\includegraphics[width=1.3in]{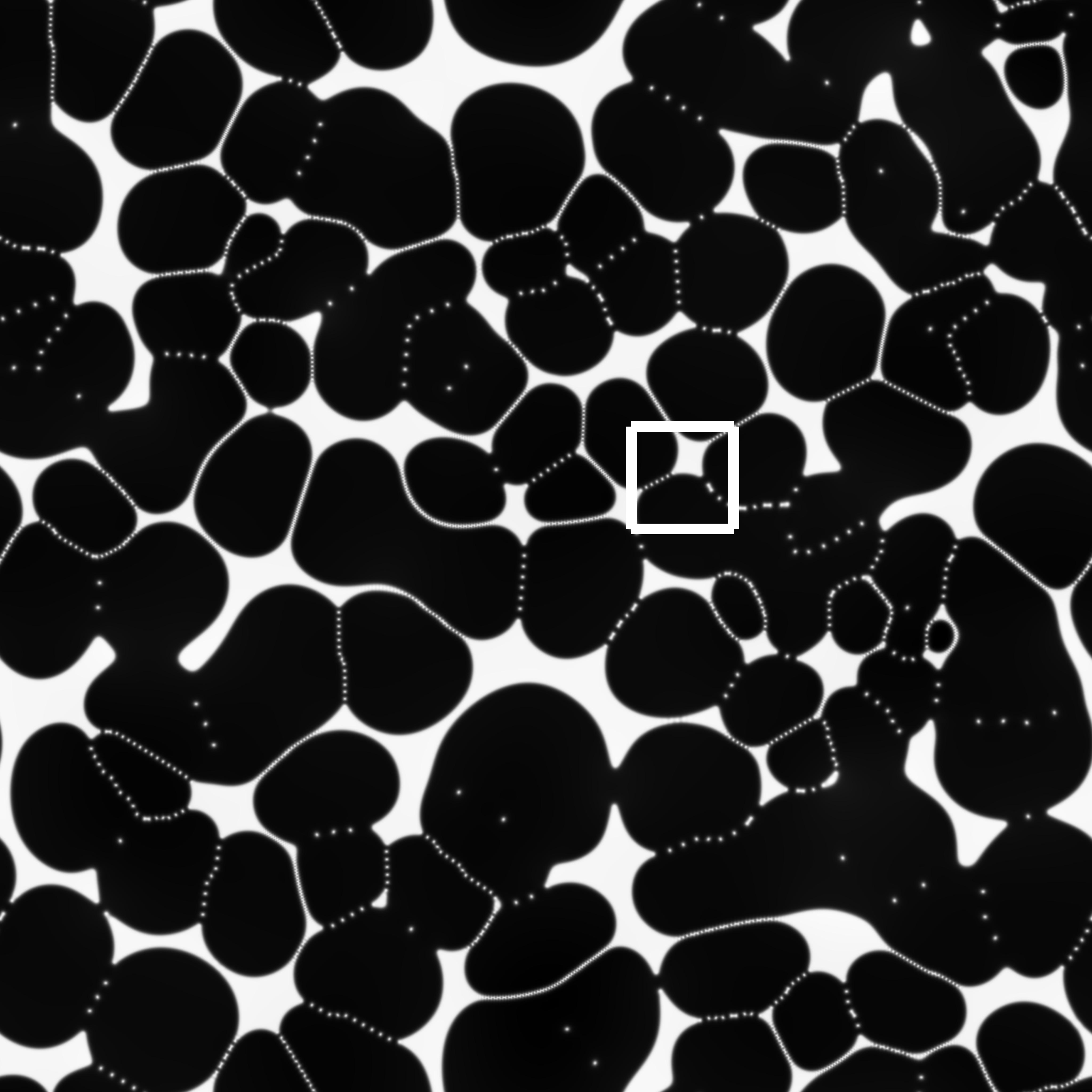}&\includegraphics[width=1.3in]{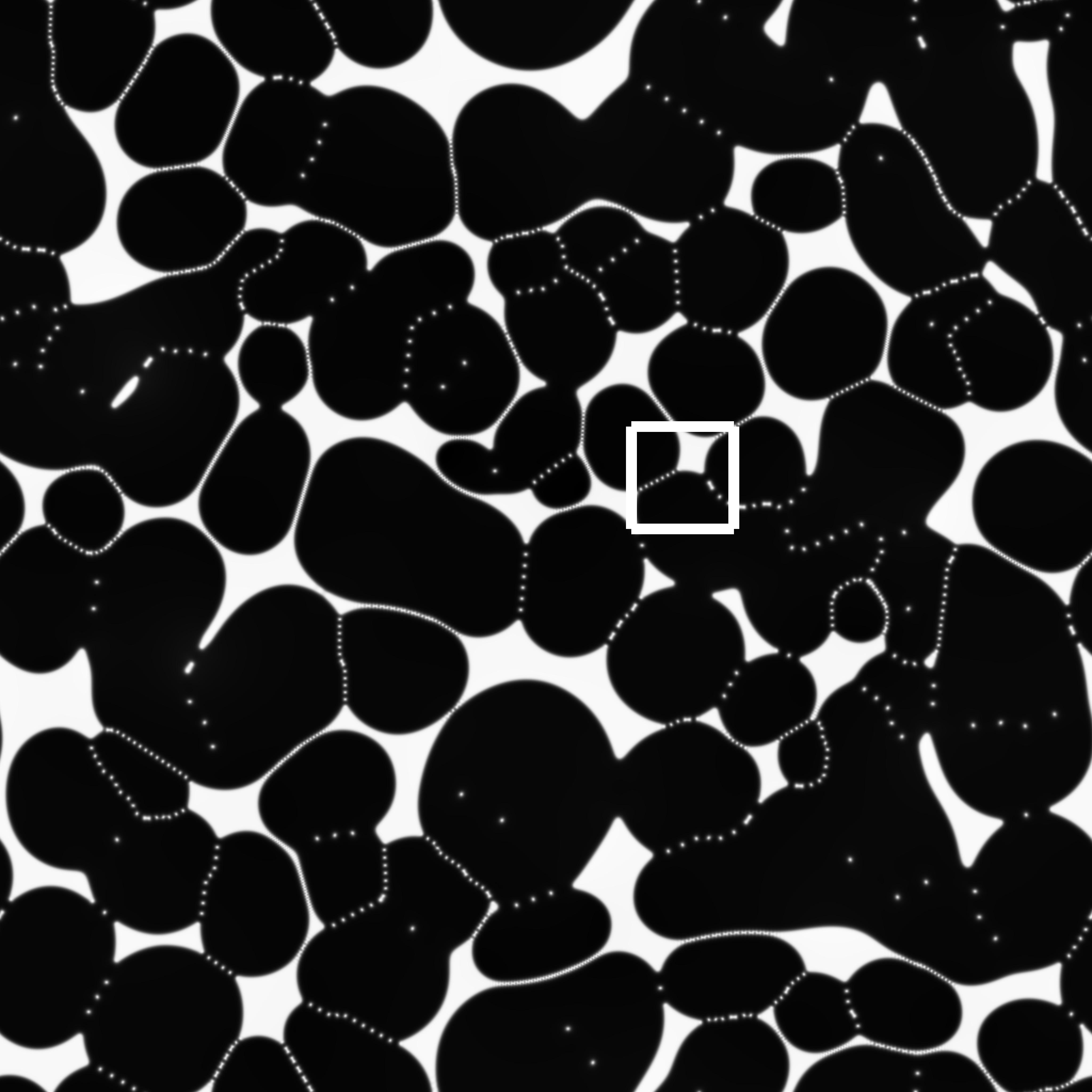}
    \\
    \includegraphics[width=1.3in]{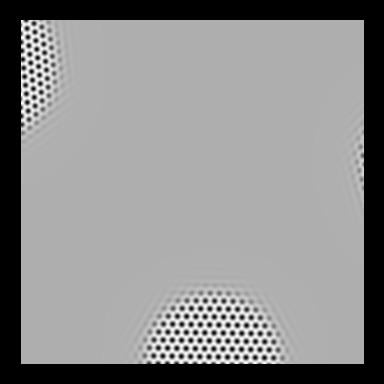}&\includegraphics[width=1.3in]{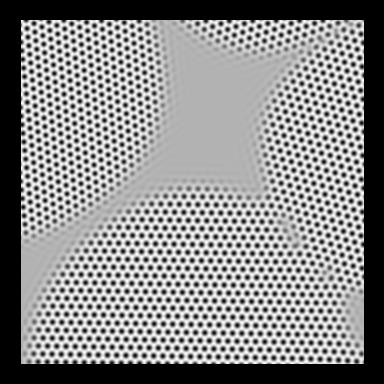}
    &\includegraphics[width=1.3in]{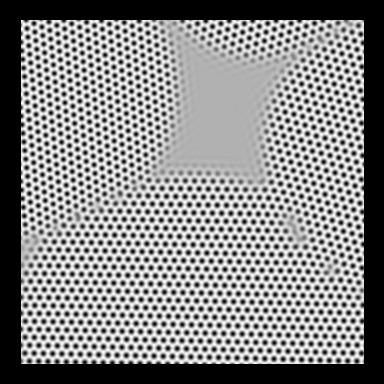}&\includegraphics[width=1.3in]{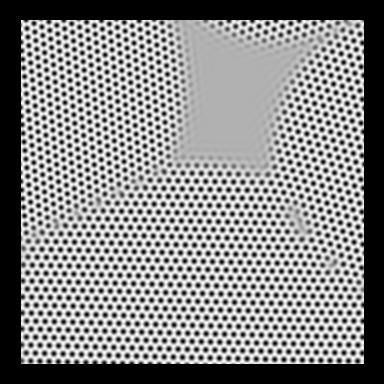}
    \\
    \includegraphics[width=1.3in]{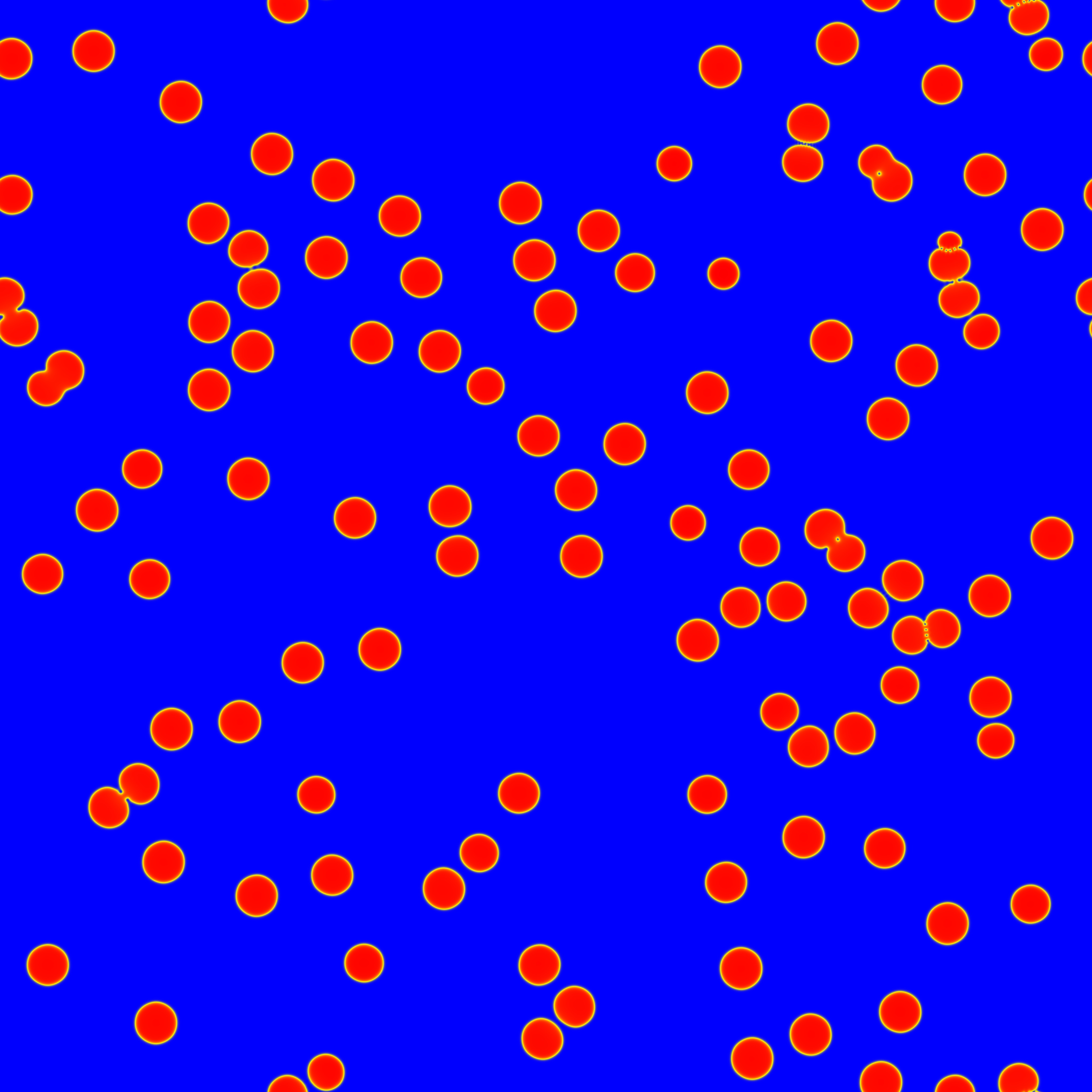}&\includegraphics[width=1.3in]{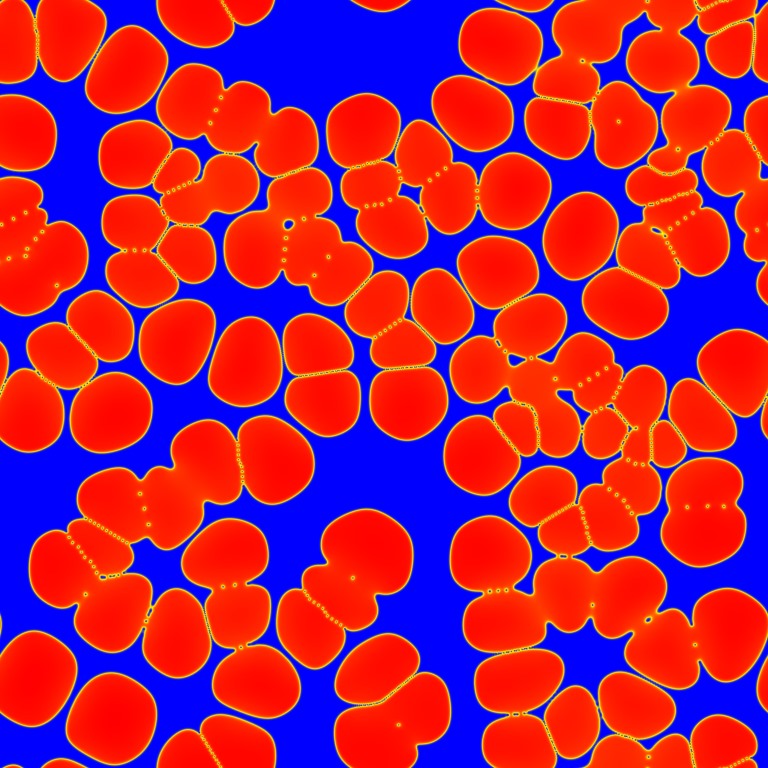}
    &\includegraphics[width=1.3in]{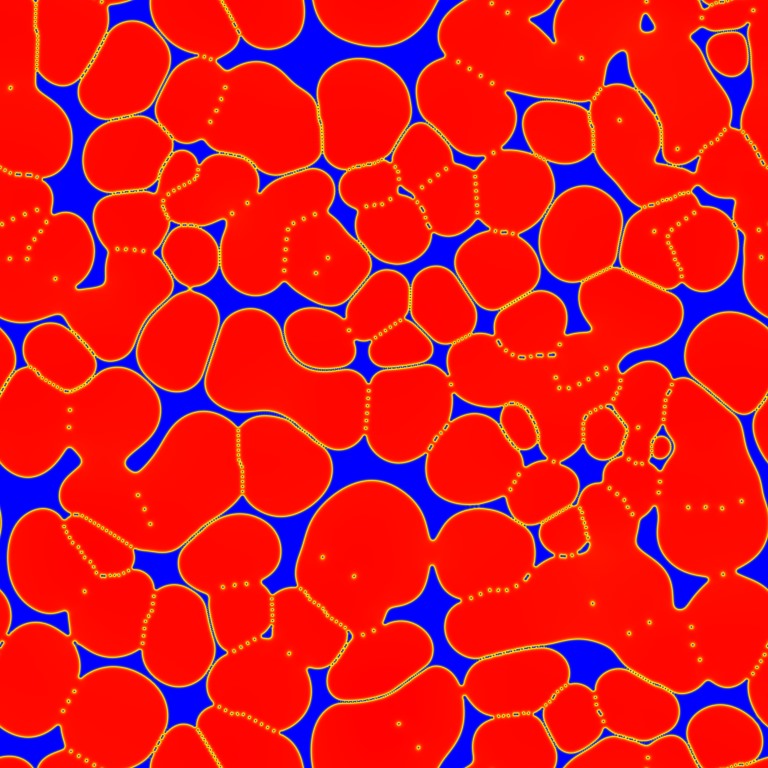}&\includegraphics[width=1.3in]{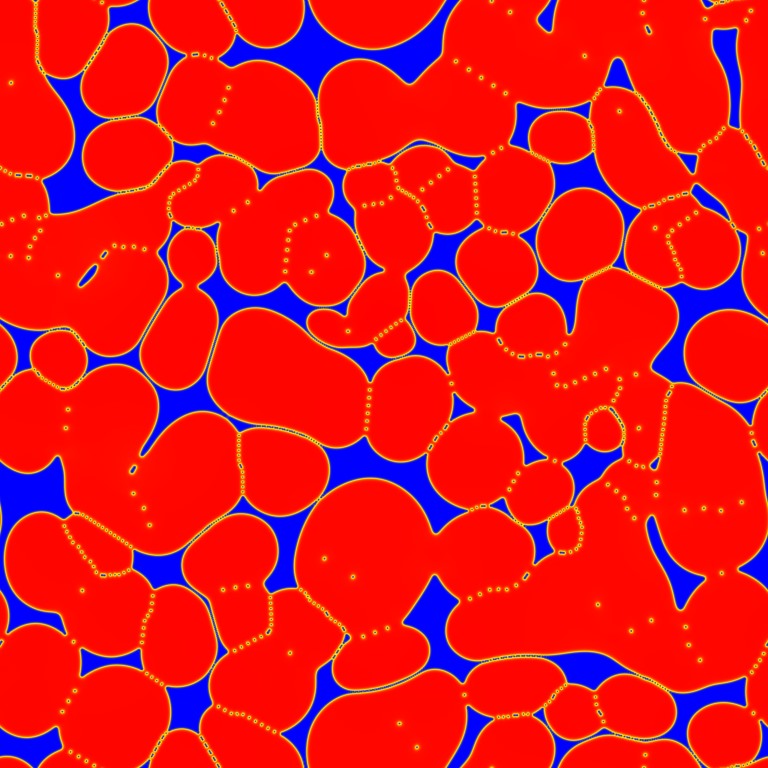}
    \end{tabular}
    \caption{(color online) Solidification and coarsening images from a simulation run. The evolution of the system progresses in time from left to right, i.e., $t=1,000, t=5,000, t=30,000$ and $t=100,000$. Top row corresponds to the average density of the grains. Large density values are black, and lower values in white. Second row represents the reconstructed atomic density of those areas marked in the top row. Probable atomic regions are darker than less probable regions and uniform values represent liquid. Third row plots the magnitude of amplitude, $A_{1}$, where red regions denote large magnitudes and blue low magnitudes.}
    \label{fig:solid-coarsening}
\end{figure*}

\section{Applications}
\label{amplitude-apps}
It is well known that most engineering materials contain multiple phases and components. While the latter is not explored in this work, we can explore a system possessing multiple phases with the amplitude formalism developed above. In this section, we demonstrate that the complex amplitude model is capable of describing two different crystal symmetries by exploring solidification, coarsening and peritectic growth. We then look at the emergence of a second phase, during grain growth, from the boundaries of a single phase polycrystalline system.

In the sections to follow, simulations were performed using Eqs.~(\ref{dynamics-no-app})-(\ref{dynamics-ampB2}). A semi-implicit Fourier technique was used to solve the system of equations. Unless stated otherwise, numerical grid spacing of $\Delta x=0.5$ and time step of $\Delta t=1$ have been used. Furthermore, all thermal fluctuations have been neglected, unless indicated otherwise. Following the original XPFC derivation, here we take $\hat{C}_2(|\mbfk=0|)=0$. For simplicity, we also take all mobility coefficients to be equal to unity, i.e., $M_{\nu}=1$, where $\nu$ is one of the corresponding fields ($n_o, \{A_j\},$ or $\{B_m\}$) in the free energy functional of Eq.~(\ref{ampEnergy}). Finally, all simulations were conducted in the phase space determined by the equilibrium phase diagram in Fig.~\ref{fig:PhaseDiagrams}.

\subsection{Solidification and coarsening}
\label{solidification-coarsening}
As a first illustration of our amplitude model, we simulate the solidification of a polycrystalline network of grains having triangular symmetry. Our simulation domain was set to $4096\times 4096$ grid spacings, which is equivalent to approximately $512\times 512$ lattice spacings. Initially, the system was seeded with $\sim 100$ triangular crystallites randomly distributed in a uniform liquid. Each crystallite had a radius of $30$ grid spacings ($\sim 4$ lattice spacings) and a randomly chosen orientation. The average density was chosen to be $n_{o}=0.115$, which at a temperature of $\sigma=0.16$ in equilibrium would give a final solid fraction of approximately $0.87$ according to the lever rule. The amplitudes of the initial triangular nuclei were chosen to satisfy $A_{j}^{\theta}=A_{j}e^{i\delta \mbfk_{j}(\theta)\cdot\mbfr}$ ($j=1,2,3$), where $A_{j}$ is the corresponding amplitudes of the original reference basis, $\delta \mbfk_{j}(\theta)=\mbfK_{j}(\theta)-\mbfk_j$, with $\theta$ being the randomly chosen orientation between the interval $[-\pi/6,\pi/6]$ and $\mbfK_{j}(\theta)$ the rotated triangular reciprocal lattice vectors.

In Fig.~\ref{fig:solid-coarsening}, we show some snapshots of the solidification and coarsening process. In descending order of rows from top to bottom, Fig.~\ref{fig:solid-coarsening} displays the average density field, the reconstructed atomic density (from a portion of the simulation domain) and the magnitude of $A_{1}$, respectively, with simulation times $t=1,000, t=5,000, t=30,000$ and $t=100,000$, increasing from left to right. As shown in the images of the average density, i.e., top row, where darker areas denote regions of high density and white regions of low, initial crystallites once nucleated, grow  and partly coalesce leading to grain boundaries ($t=5,000$), defined by the dislocations between boundaries. After the soft impingement of the grains, once the system has approximately reached the equilibrium solid fraction, we observe subsequent coarsening in frames $t=30,000$ and $t=100,000$, which occurs to minimize the total interfacial energy of the system via curvature reduction. This manifests itself in the standard process of coarsening with larger grains growing at the expense of smaller ones.

\begin{figure}[htb!]
    \centering
    \begin{tabular}{cc}
    \includegraphics[width=1.5in]{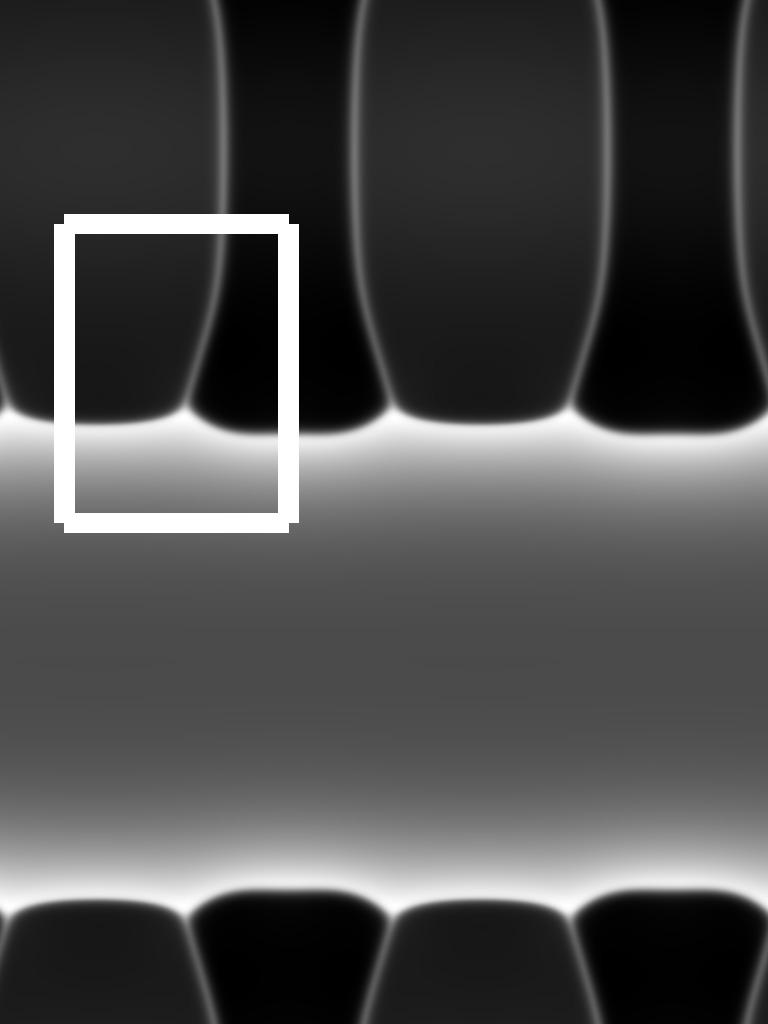}&\includegraphics[width=1.5in]{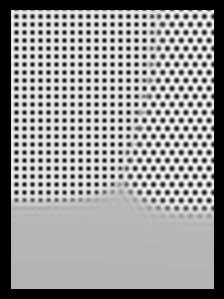}
    \\
    (a)&(b)
    \\
    \includegraphics[width=1.5in]{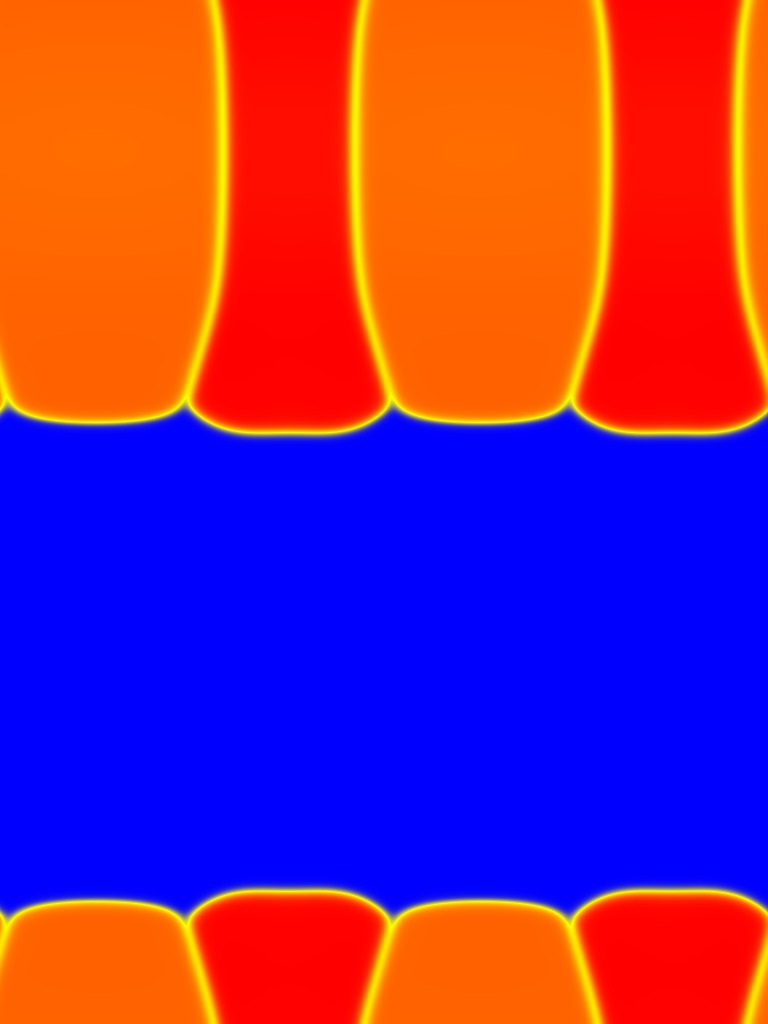}&\includegraphics[width=1.5in]{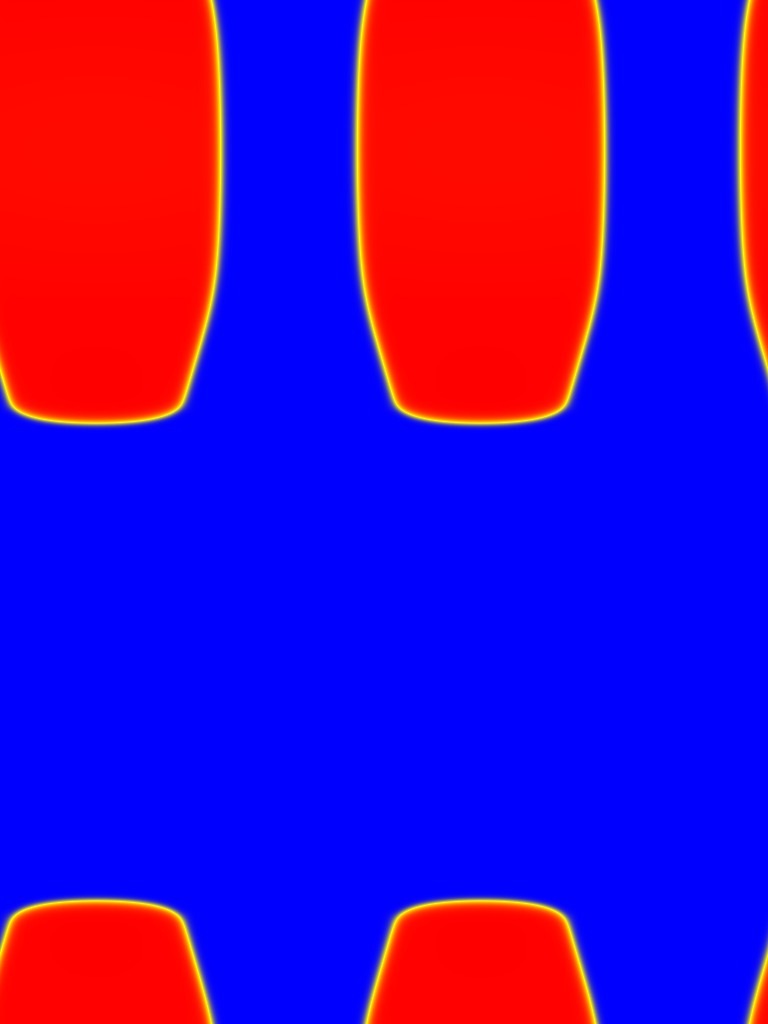}
    \\
    (c)&(d)
    \end{tabular}
    \caption{(color online) Simulation snapshots, at $t=10,000$, of peritectic solidification. (a) average density, where dark regions represent high density areas, with lighter regions low density. (b) Reconstructed atomic probability density of the marked area in (a). Areas of largest probability are darker compared to areas of lower probability. (c) Magnitude of amplitude $A_{1}$, which is nonzero in both solid structures; areas of larger magnitudes are depicted in red and zero magnitudes are blue. (d) Magnitude of amplitude $B_{1}$, which is only nonzero in the square phase. Color scheme is the same as in (c). }
    \label{fig:peritectic}
\end{figure}
\subsection{Peritectic growth}
\label{peritectic}
Our second demonstration of the above amplitude model exploits the multi-phase nature of the XPFC modeling formalism. Here we illustrate peritectic growth, where the two solid structures have different crystalline symmetries. The simulation cell was a rectangular domain of size $768\times 1024$ grid spacings ($\sim 96\times 128$ lattice spacings), where we initialized the system with alternating square and triangular structures having length $200$, and width $100$ grid spacings respectively. The average density was set to $n_{o}=0.07$, at the approximate peritectic temperature, $\sigma=0.1256$. Figure~\ref{fig:peritectic} shows a snapshot during the growth process at late time. Displayed are a selection of the various fields which make up the peritectic structure. We have in Fig.~\ref{fig:peritectic}, the average density field (top left), the reconstructed atomic density of the area marked on the average density (top right), the magnitude of  $A_1$ (non-zero for both structures, bottom left), and the magnitude of $B_1$ (non-zero for the square phase, bottom right).

\subsection{Grain growth and emergence of second phase structures}
\label{grain-growth}
To further illustrate the robust capability of the amplitude model derived in this work, here we examine the emergence of a secondary phase (square), from the grain boundaries and triple junctions of a polycrystalline network of grains having triangular symmetry. The initial condition was taken from the solidification simulation of our triangular system, in Sec.~\ref{solidification-coarsening}, at $t=5,000$. This system was quenched into the single square-phase region at a temperature of $\sigma=0.1$. The system was left for a thousand time steps to allow complete coalescence and merger of the grains having triangular symmetry. After merger, a nonzero noise amplitude of $0.001$ for all stochastic variables was introduced for all dynamic equations, thus activating thermal fluctuations in the system for one thousand time steps. Once nucleation of the square phase was apparent, the noise amplitude was set back to zero.
\begin{figure*}[htb]
    \centering
    \begin{tabular}{ccc}  \includegraphics[width=1.6in]{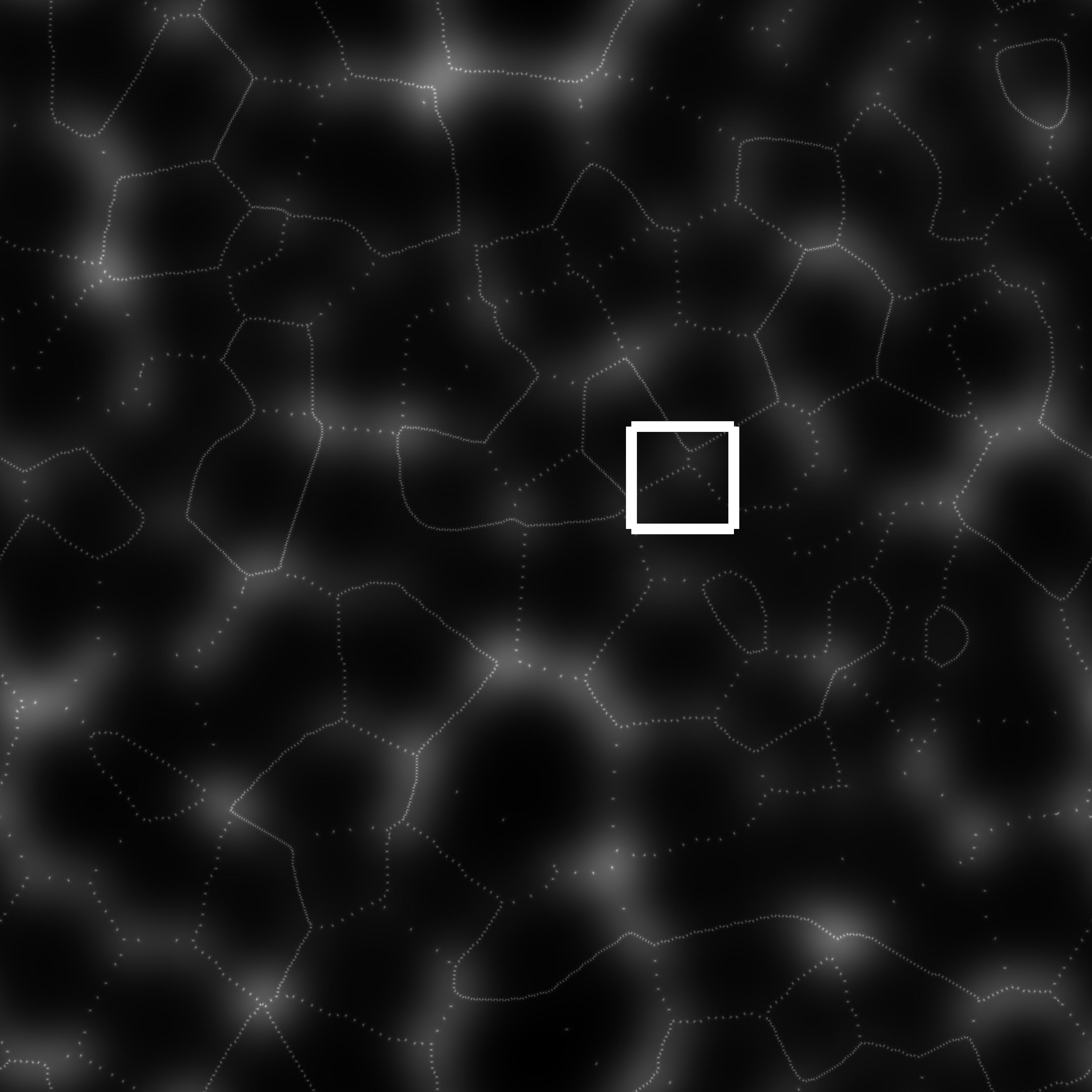}&\includegraphics[width=1.6in]{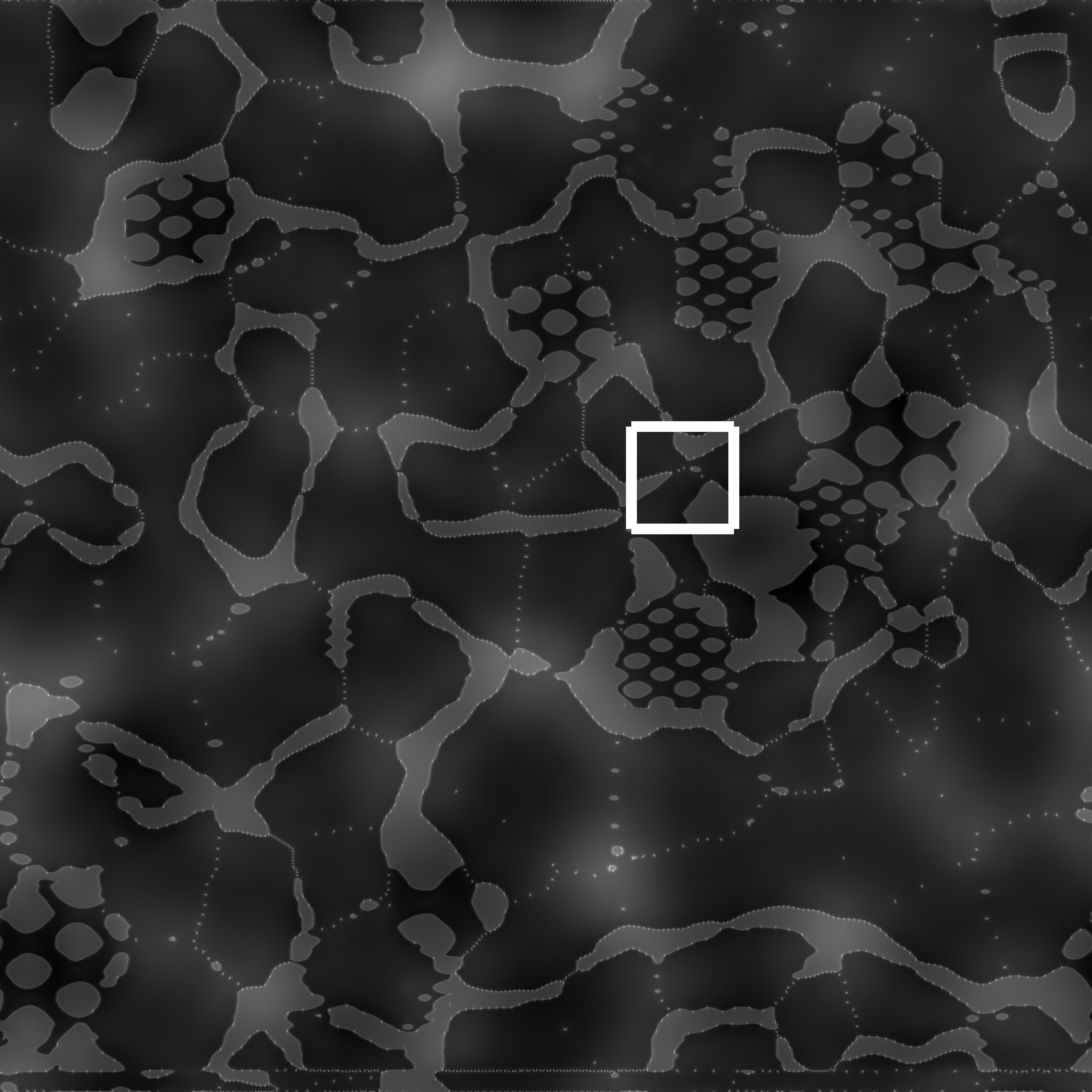}&\includegraphics[width=1.6in]{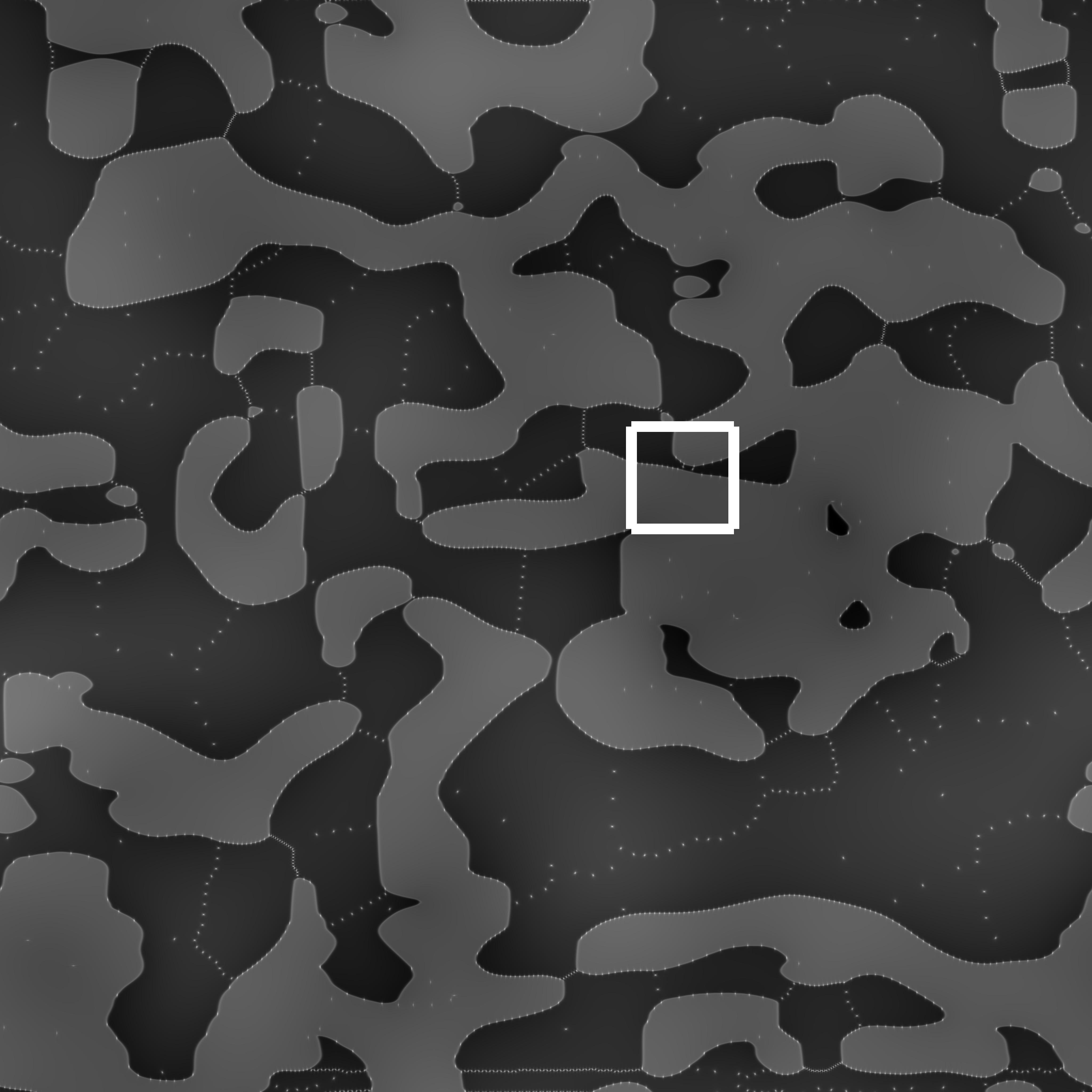}
    \\ \includegraphics[width=1.6in]{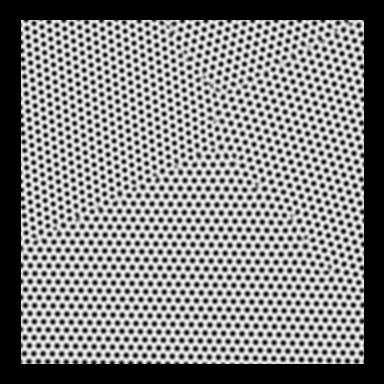}&\includegraphics[width=1.6in]{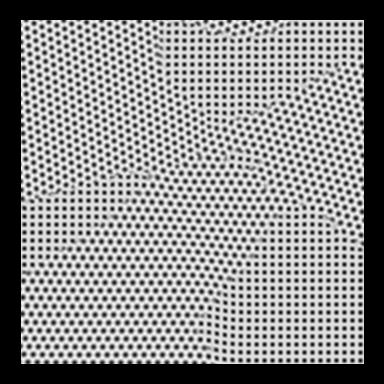}&\includegraphics[width=1.6in]{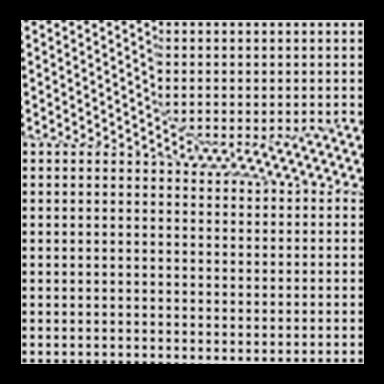}
	\\
    \includegraphics[width=1.6in]{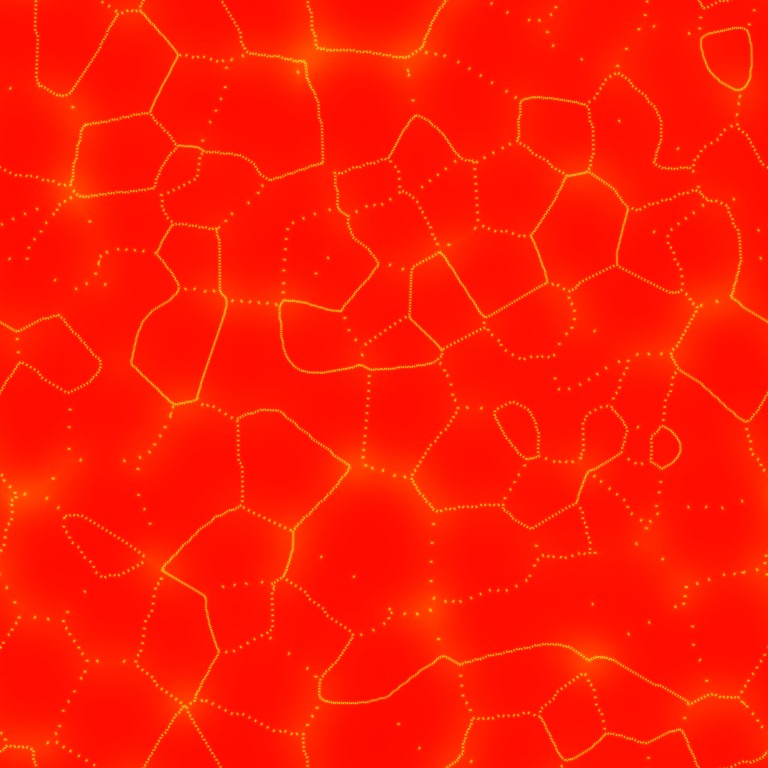}&\includegraphics[width=1.6in]{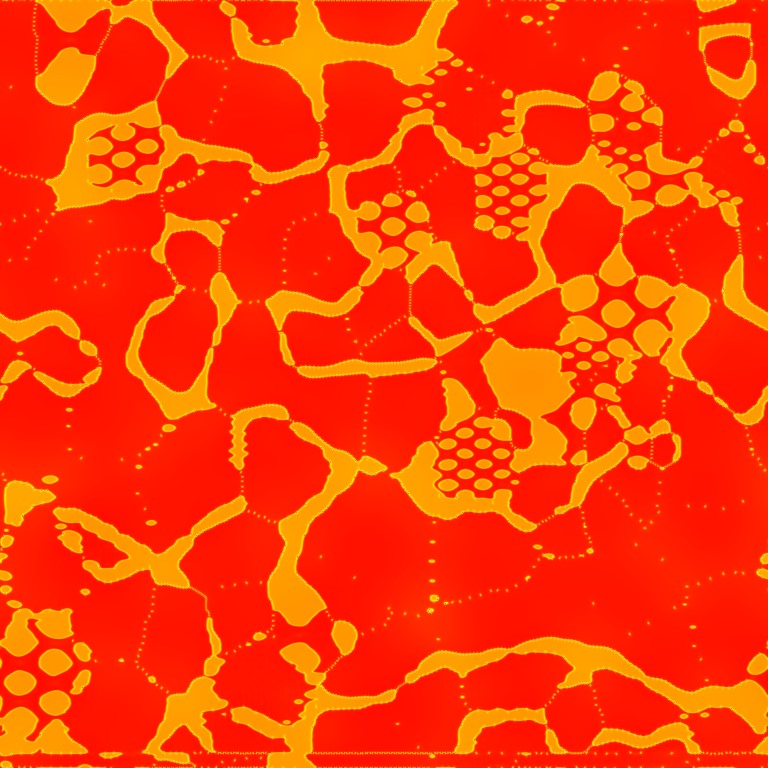}&\includegraphics[width=1.6in]{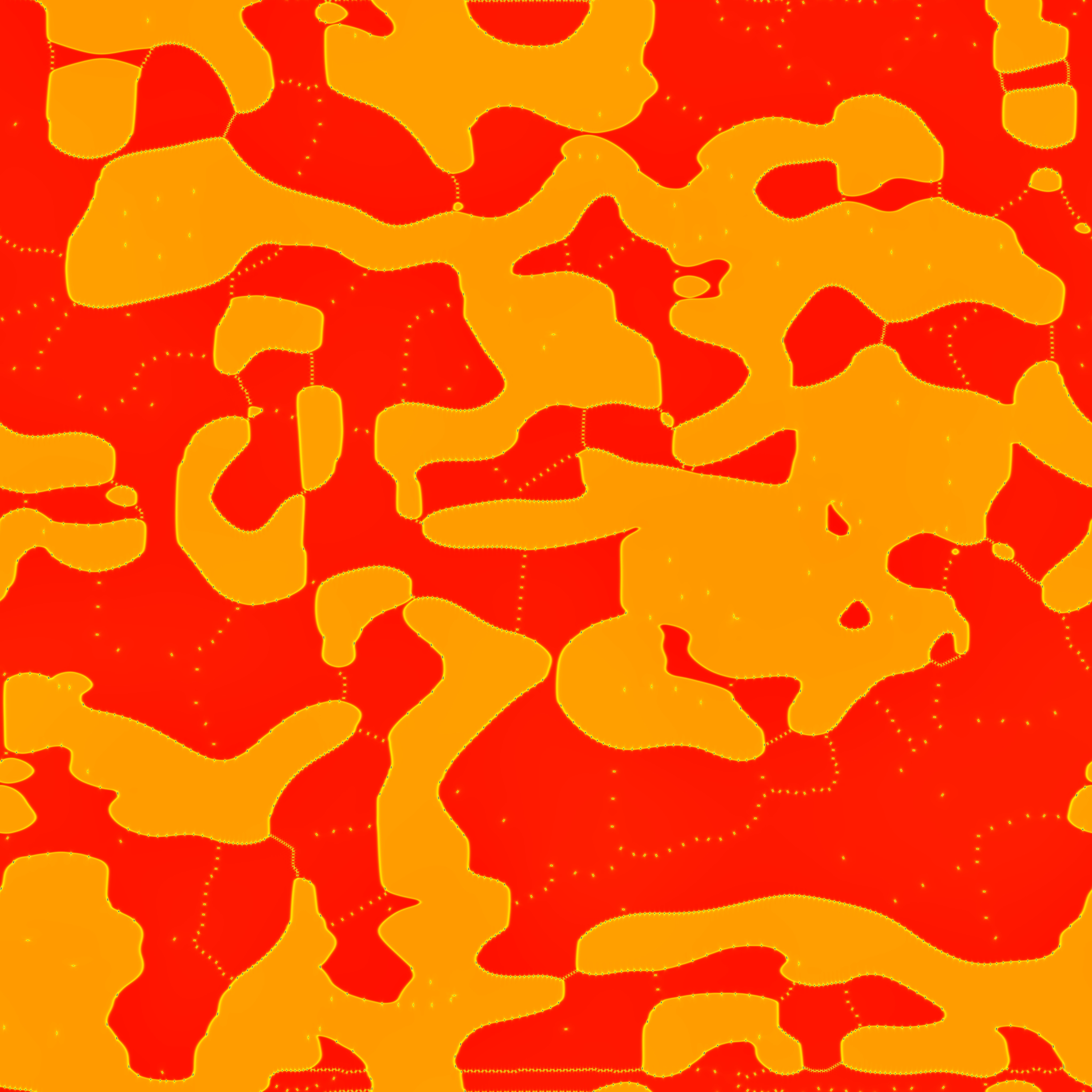}
    \\
    \includegraphics[width=1.6in]{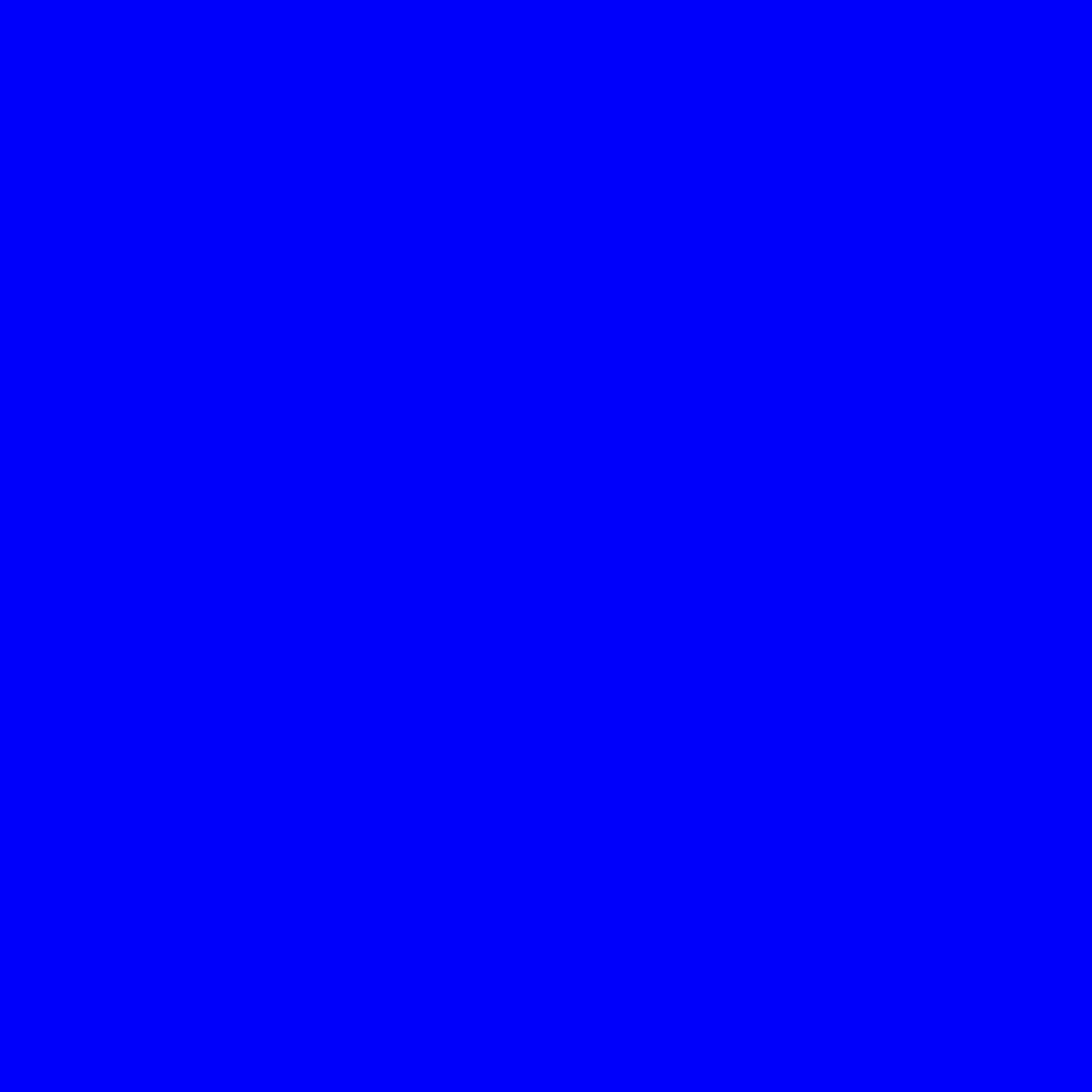}&\includegraphics[width=1.6in]{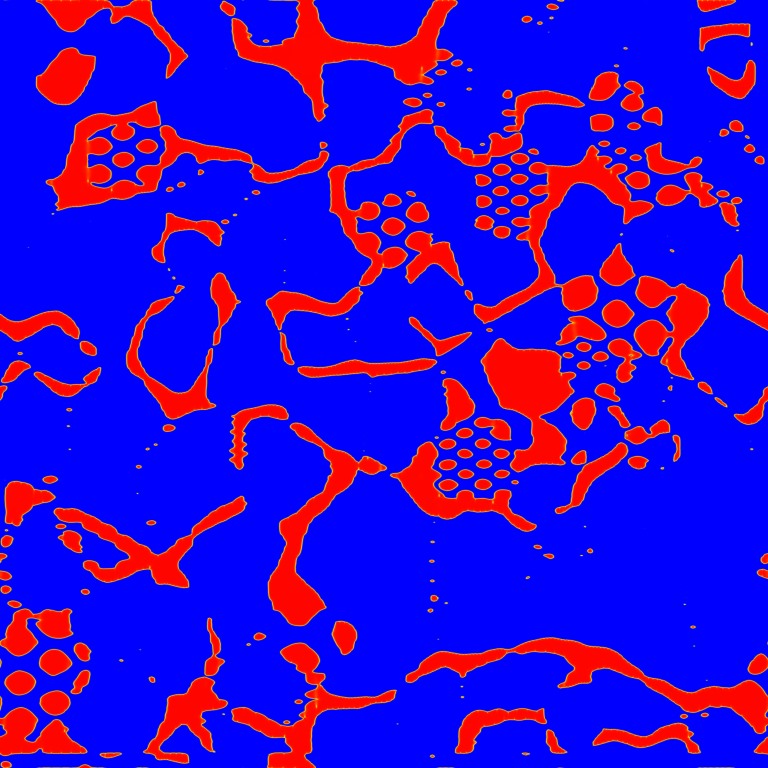}&\includegraphics[width=1.6in]{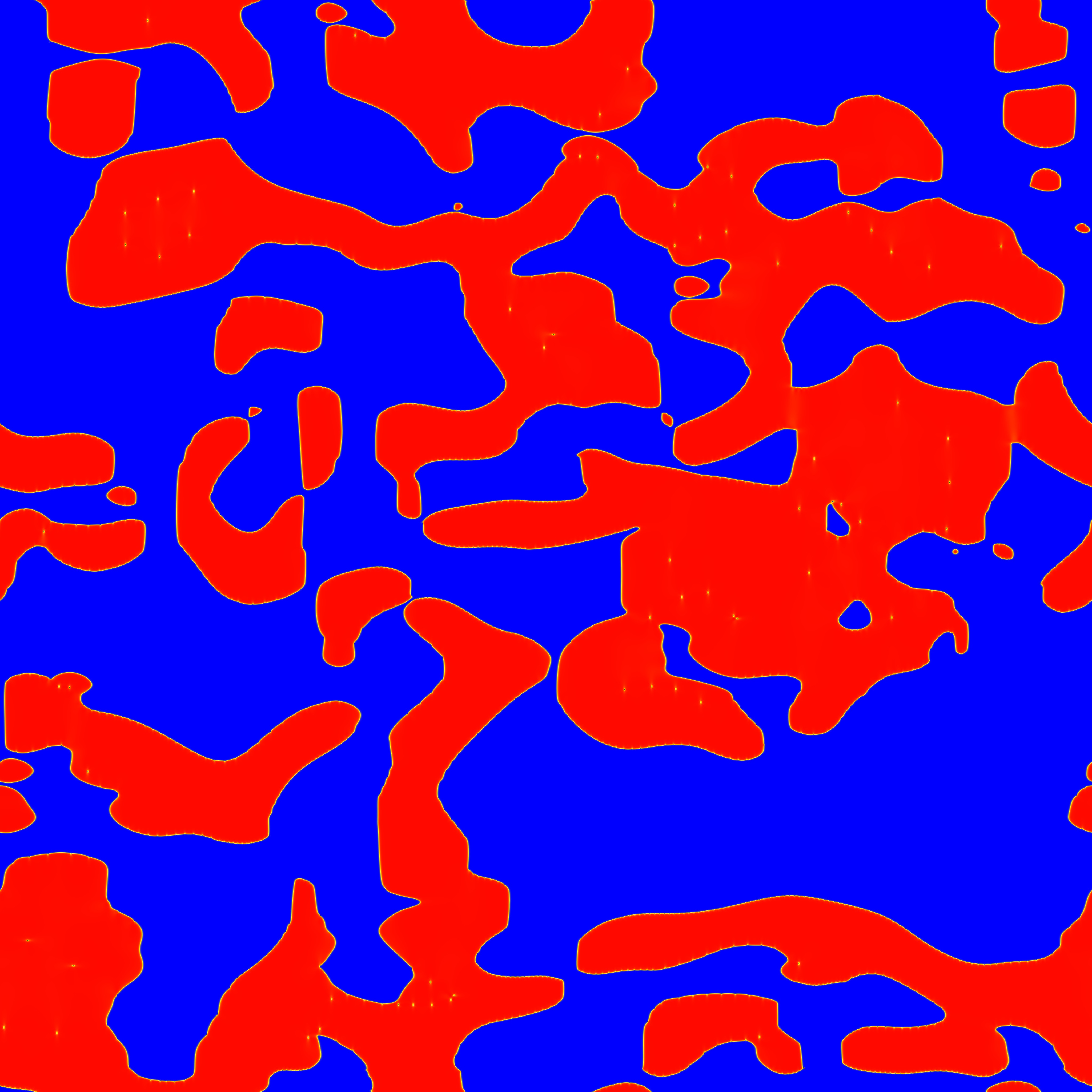}
    \end{tabular}
    \caption{(color online) Time evolution of grain growth exhibiting emergence of a secondary phase (square) at the boundaries and triple junctions of the primary solidified phase (triangular). System evolution progresses from left to right, i.e., $t=1,000$, $t=2,000$ and $t=9,000$. Top row plots the average density. Large density values are darker shades while low values are light shades. The reconstructed atomic density of the areas marked in the top row are shown in the second row. Third and fourth rows display $|A_{1}|$ and $|B_{1}|$, respectively. Red indicates areas of large magnitudes while blue represents a magnitude of zero.}
    \label{fig:grain-growth}
\end{figure*}

Figure~\ref{fig:grain-growth} shows several snapshots during the system evolution, exhibiting the emergence of the secondary square phase from the boundaries of the triangular polycrystalline network. From top to bottom, the plots display the average density, $n_{o}$, the reconstructed density, $n$, the magnitude of amplitude $A_{1}$ and the magnitude of $B_{1}$ (which is only nonzero for the square phase). Time increases from left to right in Fig.~\ref{fig:grain-growth}. Clearly evident in the progression of the images in Fig.~\ref{fig:grain-growth} is onset and subsequent growth of the secondary phase. This illustrates the further capability of our amplitude model in describing the self-consistent nucleation and growth of phases, a phenomena that cannot be captured currently with phase-field and other mean field type formalisms.

\section{Summary}
\label{summary}
In this paper, we reported on a new Fourier technique for deriving complex amplitude models for PFC and PFC-type free energy functionals. Details of the method were discussed in the context of the structural PFC formalism for single component systems in 2D. Our approach was also shown to recover forms of previous approaches, as well as address the issue of the periodic instability of the average density. The dynamics of the model were demonstrated with simulations of solidification and coarsening, peritectic solidification involving different crystal structures, and grain growth exhibiting nucleation and growth of  a secondary phase, phenomena of relevance in microstructural evolution, where the latter two cannot be captured with currently available mean field formalisms such as the phase-field method.

Complex amplitude models were introduced as a way to provide a link between the standard phase-field approach and the phase-field-crystal approach. Having developed a complex amplitude model capable of describing multiple crystal structures and elasto-plastic effects, this work has demonstrated the nature of such a bridge between the methodologies by directly incorporating the properties of the microscopic correlation function. Operating on larger scales, the model was shown to capture the salient atomistic scale features inherent in several important phase transformations currently outside the capability of the standard phase-field approach. As a novel technique, our method can accept as input any derived or experimentally calculated correlation function, which makes it applicable to a myriad of systems. It is expected that such a method, when combined with novel mesh algorithms, can truly represent a multi-scale modeling paradigm for investigating microstructural processes governed by elasticity and defects operating on diffusional time and length scales.

\acknowledgements{We thank Sami Majaneimi, Michael Greenwood and Ken Elder for their insight and constructive discussions. N.P. acknowledges support from the National Science and Engineering Research Council of Canada (NSERC). Z.F.H. acknowledges support from the National Science Foundation under Grant No. DMR-0845264. We also thank Compute Canada, particularly Clumeq and Sharcnet, for computing resources.}

\appendix
\section{Dynamic Equations}
\label{dynamicsFull}
In Sec.~\ref{dynamics}, we introduced the variational principles applied to the coarse-grained free energy functional, $F^{cg}$, in arriving at the set of dynamic equations. Here, we explicitly apply the variational principles and write the resulting equations of motion.

For the average density we have,

\begin{align}
\label{dynamics-no-app}
\frac{\partial n_o}{\partial t} \!\! &= \!\nabla \! \cdot \! \Bigg( M_{n_o}\nabla \Bigg\{n_o - \eta\frac{n_o^2}{2}+\chi\frac{n_o^3}{3} -\lt[\hat{\xi}_{V}(\mbfk)\hat{C}_{2}(\mbfk)\hat{n}_{o}(\mbfk)\rt]_{\mbfr}\nonumber\\
 &+\lt(2\chi\,n_o-\eta\rt)\bigg(\sum_{j}^4|A_j|^2 + \sum_{m}^2|B_m|^2\bigg)\nonumber\\
& +2\chi\lt[A_1A_2A_3 + A_1^*A_4^*B_1 + A_1A_4^*B_2^* + c.c. \rt]\Bigg\}\Bigg).
\end{align}

Equations for the first mode of the amplitudes read,
\begin{align}
\label{dynamics-ampA1}
\frac{\partial A_{1}}{\partial t} \!\! &= -M_{A_{1}}\Bigg\{\lt(1-\eta\,n_o+ \chi\,n^2_o\rt)A_1\nonumber\\
&-(\eta-2\chi n_o)\lt[A_2^*A_3^* + A_4B_2 + A_4^*B_1\rt]\nonumber\\
&+\chi A_1\lt(|A_1| + 2\lt[\sum_{j\ne1}^{4}|A_j|+\sum_{m}^{2}|B_m|\rt]\rt)\nonumber\\
&+2\chi 2A_1^*B_1B_2-\lt[\hat{C}_2(|\mbfk+\mbfk_1|)\hat{A}_{1}(\mbfk)\rt]_{\mbfr}\Bigg\},
\end{align}
\begin{align}
\label{dynamics-ampA2}
\frac{\partial A_{2}}{\partial t} \!\! &= -M_{A_{2}}\Bigg\{\lt(1-\eta\,n_o+ \chi\,n^2_o\rt)A_2\nonumber\\
&-(\eta-2\chi n_o)A_1^{*}A_3^{*}+2\chi A_3^{*}\lt[A_4B_1^{*}+A_4^{*}B_2^{*}\rt]\nonumber\\
&+\chi A_2\lt(|A_2| + 2\lt[\sum_{j\ne2}^{4}|A_j|+\sum_{m}^{2}|B_m|\rt]\rt)\nonumber\\
&-\lt[\hat{C}_2(|\mbfk+\mbfk_2|)\hat{A}_{2}(\mbfk)\rt]_{\mbfr}\Bigg\},
\end{align}
\begin{align}
\label{dynamics-ampA3}
\frac{\partial A_{3}}{\partial t} \!\! &= -M_{A_{3}}\Bigg\{\lt(1-\eta\,n_o+ \chi\,n^2_o\rt)A_3\nonumber\\
&-(\eta-2\chi n_o)A_1^{*}A_2^{*}+2\chi A_2^{*}\lt[A_4B_1^{*}+A_4^{*}B_2^{*}\rt]\nonumber\\
&+\chi A_3\lt(|A_3| + 2\lt[\sum_{j\ne3}^{4}|A_j|+\sum_{m}^{2}|B_m|\rt]\rt)\nonumber\\
&-\lt[\hat{C}_2(|\mbfk+\mbfk_3|)\hat{A}_{3}(\mbfk)\rt]_{\mbfr}\Bigg\},
\end{align}
\begin{align}
\label{dynamics-ampA4}
\frac{\partial A_{4}}{\partial t} \!\! &= -M_{A_{4}}\Bigg\{\lt(1-\eta\,n_o+ \chi\,n^2_o\rt)A_4\nonumber\\
&-(\eta-2\chi n_o)\lt[A_1^{*}B_1+A_1B_2^{*}\rt]\nonumber\\
&+\chi A_4\lt(|A_4| + 2\lt[\sum_{j\ne4}^{4}|A_j|+\sum_{m}^{2}|B_m|\rt]\rt)\nonumber\\
&+\chi\lt[2A_2A_3B_1 + 2A_4^*B_1B_2^* + 2A_2^*A_3^*B_2^*\rt]\nonumber\\
&-\lt[\hat{C}_2(|\mbfk+\mbfk_4|)\hat{A}_{4}(\mbfk)\rt]_{\mbfr}\Bigg\}.
\end{align}

Finally, for the second set of amplitudes, corresponding to the second set of reciprocal lattice vectors, we have
\begin{align}
\label{dynamics-ampB1}
\frac{\partial B_{1}}{\partial t} \!\! &= -M_{B_{1}}\Bigg\{\lt(1-\eta\,n_o+ \chi\,n^2_o\rt)B_1-(\eta-2\chi n_o)A_1A_4\nonumber\\
&+\chi B_1\lt(|B_1| + 2\lt[\sum_{j}^{4}|A_j|+|B_2|\rt]\rt)\nonumber\\
&+\chi\lt[2A_4A_2^{*}A_3^{*} + A_4^{2}B_2 + A_1^{2}B_2^{*}\rt]\nonumber\\
&-\lt[\hat{C}_2(|\mbfk+\mbfq_1|)\hat{B}_{1}(\mbfk)\rt]_{\mbfr}\Bigg\},
\end{align}
\begin{align}
\label{dynamics-ampB2}
\frac{\partial B_{2}}{\partial t} \!\! &= -M_{B_{2}}\Bigg\{\lt(1-\eta\,n_o+ \chi\,n^2_o\rt)B_2-(\eta-2\chi n_o)A_1A_4^{*}\nonumber\\
&+\chi B_2\lt(|B_2| + 2\lt[\sum_{j}^{4}|A_j|+|B_1|\rt]\rt)\nonumber\\
&+\chi\lt[2A_2^{*}A_3^{*}A_4^{*} + (A_4^{*})^{2}B_1 + A_1^{2}B_1^{*}\rt]\nonumber\\
&-\lt[\hat{C}_2(|\mbfk+\mbfq_2|)\hat{B}_{2}(\mbfk)\rt]_{\mbfr}\Bigg\}.
\end{align}

\section{Amplitude Equations for 12 Vector Density Expansion}
\label{energy12vector}
In Sec.~\ref{dens-expan}, where we considered a density mode expansion for our coarse-graining procedure, we arrived at two expansions. While in the text, we opted to go with the simpler of the expansions, it was not motivated from any physical arguments or considerations, but rather for convenience. In this appendix, we present the coarse-grained free energy functional associated with the density mode expansion containing 12 complex amplitudes. Before proceeding, we recall the density expansion of the form
\beq
n(\mbfr) = n_o(\mbfr) + \sum_j^{6} A_j(\mbfr)e^{i\mbfk_j\cdot\mbfr} + \sum_m^{6} B_m(\mbfr)e^{i\mbfq_m\cdot\mbfr} + c.c.
\label{densExpan12b}
\eeq
The derivation of the amplitude equation for $12$ amplitudes is motivated and follows from the same arguments and approximations that lead us to the coarse-grained free energy functional of the simpler 6 complex amplitude energy of Eq.~(\ref{ampEnergy}). The coarse-grained free energy functional of the 12 complex amplitude expansion reads,
\begin{widetext}
\begin{align}
F_{12}^{cg} &= \int d\mbfr \Bigg\{\frac{n_o^2}{2} - \eta\frac{n_o^3}{6} + \chi\frac{n_o^4}{12} + \lt(1-\eta\,n_o+ \chi\,n^2_o\rt)\,\bigg(\sum_{j}^{6}|A_j|^2 + \sum_{m}^{6}|B_m|^2\bigg) \nline
&-(\eta-2\chi n_o)\Big[A_1A_2A_3 + A_4A_5A_6 + B_1B_2B_3 + A_2A^*_5B_5 + A_4A^*_1B_4 \nonumber \\
&+ A_3A_6B^*_3 + A_3A^*_6B_6 + A_2A_5B^*_2 + A_1A_4B^*_1 + B_5B_6B^*_4 + c.c.\Big]\nline
&+\frac{\chi}{2}\lt[\sum_{j}^{6}A_j^2(A^*_j)^2 +  \sum_{m}^{6}B_m^2(B^*_m)^2\rt]+2\chi \lt[\sum_j^{6}\sum_{m>j}^{6}|A_j|^2|A_m|^2 + \sum_{j}^{6}\sum_{m}^{6}|A_j|^2|B_m|^2 + \sum_j^{6}\sum_{m>j}^{6}|B_j|^2|B_m|^2\rt] \nline
&+\chi\Big[ A_1^2B_1^*B_4^* + A_2^2B_5B_2^* +A_3^2B_6B_3^* + A_4^2B_4B_1^* + A_5^2B_2^*B_5^* + A_6^2B_3^*B_6^* + c.c.\Big]\nonumber \\
&+2\chi\Big[ A_2A_3A_4B_4 + A_2A_4A_6B_5 + A_3A_4A_5B_6 + A_1A_2A_6B_6^* + A_1A_3A_5B_5^* + A_1A_5A_6B_4^*\nonumber \\
&+ A_1A_3A_5^*B_2 + A_1A_2A_6^*B_3 + A_2A_3A_4^*B_1 + A_4A_5A_3^*B_3 + A_4A_6A_2^*B_2 + A_5A_6A_1^*B_1\nonumber\\
&+ A_1A_4B_2B_3 + A_2A_5B_1B_3 + A_3A_6B_1B_2 + A_4A_1^*B_5B_6 + A_3A_6^*B_4B_5^* + A_5A_2^*B_6B_4^* + c.c.\Big]\nonumber\\
&-\frac{n_o}{2}\lt[\hat{\xi}_{V}(\mbfk)\hat{C}_2(|\mbfk|)\hat{n}_{o}(\mbfk)\rt]_{\mbfr}-\frac{1}{2}\sum_{j}^{6}A_{j}^{*}\lt[\hat{C}_2(|\mbfk+\mbfk_j|)\hat{A}_{j}(\mbfk)\rt]_{\mbfr} -\frac{1}{2}\sum_{j}^{6} A_{j}\lt[\hat{C}_2(|\mbfk-\mbfk_j|)\hat{A}_{j}(-\mbfk)\rt]_{\mbfr} \nonumber \\
&-\frac{1}{2}\sum_{m}^{6}B_{m}^{*}\lt[\hat{C}_2(|\mbfk+\mbfq_m|)\hat{B}_{m}(\mbfk)\rt]_{\mbfr} -\frac{1}{2}\sum_{m}^{6} B_{m}\lt[\hat{C}_2(|\mbfk-\mbfq_m|)\hat{B}_{m}(-\mbfk)\rt]_{\mbfr}\Bigg\}.
\label{12ampEnergy}
\end{align}
\end{widetext}

%
\end{document}